\documentclass[12pt,english,sort&compress, letterpaper, ]{JHEP3}
\usepackage[T1]{fontenc}
\usepackage{ae}
\usepackage{aecompl}
\usepackage[latin9]{inputenc}
\usepackage{amsthm}
\usepackage{amsmath}
\usepackage{amssymb}
\usepackage[numbers]{natbib}

\makeatletter
\title{Hypermoduli Stabilization, Flux Attractors, and Generating Functions}
\preprint{MCTP-09-60 \\ MIFP-09-53}
\author{Finn Larsen and Ross O'Connell \\ 
Department of Physics, University of Michigan,\\
450 Church Street, Ann Arbor, MI 48109-1020, USA \\ 
E-mail: \email{larsenf@umich.edu}, \email{rcoconne@umich.edu}}
\author{Daniel Robbins\\
George P. and Cynthia W. Mitchell Institute for Fundamental Physics,\\
Texas A \& M University, College Station, TX 77843, USA\\
E-mail: \email{robbins@physics.tamu.edu}}
\abstract{We study stabilization of hypermoduli with emphasis on the effects of
generalized fluxes. We find a class of no-scale vacua described by ISD conditions 
even in the presence of geometric flux. The associated flux attractor equations 
can be integrated by a generating function with the property that the hypermoduli
are determined by a simple extremization principle. We work out several orbifold
examples where all vector moduli and many hypermoduli are stabilized, with VEVs given explicitly in terms of 
fluxes.}

\makeatother

\usepackage{babel}

\begin{document}
\global\long\def\k{\kappa}
 \global\long\def\al{\alpha}
 \global\long\def\m{\mu}
 \global\long\def\whk{\widehat{\kappa}}
 \global\long\def\w{\wedge}
\global\long\def\non{\,}
\global\long\def\Om{\Omega}
\global\long\def\g{\gamma}
 \global\long\def\om{\omega}
 \global\long\def\hlf{\frac{1}{2}}
\global\long\def\N{\mathcal{N}}
\global\long\def\Z{\mathbb{Z}}
\global\long\def\R{\mathbb{R}}
\global\long\def\p{\partial}
\global\long\def\t{\theta}
\global\long\def\s{\sigma}

\section{Introduction}

Many closed-string backgrounds with 4D $\mathcal{N}=1$ supersymmetry
descend from backgrounds with 4D $\mathcal{N}=2$ supersymmetry. The
chiral multiplets in the $\mathcal{N}=1$ theories then arise from
projections of either $\mathcal{N}=2$ vector multiplets or $\mathcal{N}=2$
hypermultiplets. While it is well-understood how to use fluxes to
stabilize the vector moduli in both IIB and IIA compactifications
\citep{Giddings:2001yu,DeWolfe:2005uu,Larsen:2009fw} (for review
see \citep{Grana:2005jc,Douglas:2006es,Denef:2008wq}), it has been
less clear how to stabilize the hypermoduli. In this paper we introduce
a general scheme for understanding how specific geometric and non-geometric
fluxes can stabilize many more of the hypermoduli. As an example of
this scheme, we will study in detail the addition of geometric fluxes
to O3/O7 compactifications with 3-form flux.

Our starting point is fairly general. We make only two modest assumptions
about how the hypermultiplet moduli enter into the 4D scalar potential: 
\begin{itemize}
\item \textit{Homogeneity}: the Kähler potential is assumed homogeneous
of degree four in the imaginary parts of the hypermoduli. This is
satisfied by Calabi-Yau orientifolds, as well as for more general
compactifications with $SU\left(3\right)\times SU\left(3\right)$
structure. 
\item \textit{Linearity}: the hypermultiplet moduli should only appear linearly
in the superpotential. This is the case for compactifications with
generalized fluxes, our main example. 
\end{itemize}
We show that under these assumptions the scalar potential can be rewritten
as a sum of universal, positive semi-definite terms, and a term governed
by a metric that generally has indefinite signature. For some choices
of fluxes, this final term is \emph{also} positive semi-definite,
so that we can find absolute minima of the scalar potential by setting
each term separately to zero. The resulting Minkowski vacua are natural
generalizations of the familiar no-scale vacua introduced by GKP \citep{Giddings:2001yu}.

Our primary example of this new class of Minkowski vacua is IIB O3/O7
compactifications with 3-form fluxes (as usual) and geometric fluxes
(the new ingredient). Minimization of the scalar potential for this
entire class of vacua is equivalent to an ISD condition with additional
constraints. The ISD condition can be recast as a set of flux attractor
equations~\citep{Larsen:2009fw,Kallosh:2005ax,Bellucci:2007ds,Anguelova:2008fm,Anguelova:2009az,Cassani:2009na}
which stabilize the vector moduli $z^{i}$ in the manner previously
studied for O3/O7 compactifications with 3-form fluxes alone. The
hypermoduli $\tau$ (the axio-dilaton) and $G^{a}$ (additional 2-form
moduli) enter as fixed background parameters for the purpose of vector
moduli stabilization. However, as we mentioned already, the ISD condition
is supplemented by additional constraints. It is those that control
the stabilization of the hypermoduli.

An important ingredient in the analysis is the manner in which the
vector moduli are stabilized: solutions to the flux attractor equations
can be presented as derivatives of a scalar \textit{generating function}
\citep{Larsen:2009fw}, whether or not there are geometric fluxes
present. The hypermoduli enter this generating function as parameters
that are arbitrary \textit{a priori}. However, the constraints that
control the stabilization of hypermoduli turn out to be equivalent
to an \textit{extremization principle} on the generating function
over hypermoduli space. For the purpose of hypermoduli stabilization
the generating function thus plays a role similar to that of a conventional
potential.

In favorable circumstances the extremization over hypermoduli space
may yield hypermoduli that are all stabilized at finite values. However,
it may also give either runaway behavior, or flat directions. We will
see choices of fluxes which realize each of these possibilities. An
obvious general rule is that vacua with many fluxes turned on have
fewer unstabilized moduli. More surprisingly, we find that the number
of hypermoduli that can be stabilized is apparently limited by the
number of vector moduli.

One of the motivations for this work is to develop the generating
function formalism for flux compactifications. Certainly the generating
function provides a convenient way to summarize the VEVs and the masses
of the scalar fields stabilized by fluxes. Additionally, it is intriguing
that the role it plays in the flux attractor equations is analogous
to that played by the black hole entropy in black hole attractor equations.
This analogy suggests a deep relation to counting of vacua which is
obscured by the usual geometric treatment of the fluxes. It would
be interesting to develop this relation further.

This paper is organized as follows. In section 2 we review a few features
of no-scale vacua, as they appear in the standard GKP context. We
then generalize those constructions to the generic setup of interest
here. In section 3 we provide a brief introduction to generalized
fluxes and the manner in which they appear in the low energy theory.
Combining with the results from section 2, we find the attractor equations
for no-scale vacua with geometric fluxes. In section 4 we introduce
the generating function and show that it both solves the attractor
equations for vector moduli and also provides an extremization principle
on hypermoduli space. In section 5 we give several explicit examples
that illustrate our methods. A few technical details have been collected
in appendix A.

\section{\label{sec:ScalarPotential}A General Class of no-scale Vacua}

In this section we seek to present the scalar potential of Type II
flux compactifications as a BPS-like sum of positive semi-definite
terms, thus finding Minkowski vacua when each of those terms
vanish separately. The resulting minimization conditions will later be the linchpin
for attractor equations.

We first review the standard GKP flux vacua with just one hypermodulus,
the axio-dilaton $\tau$, and then generalize to situations with many
hypermoduli. We maintain a rather general setting, albeit with assumptions
on the theory motivated by subsequent applications to situations with
generalized fluxes.

\subsection{\label{sub:O3/O7}GKP Compactifications}

To get started we review the simplest and most widely studied class
of flux vacua \citep{Dasgupta:1999ss,Giddings:2001yu}: O3/O7 orientifold
vacua in type IIB theory, with $F_{3}$ and $H_{3}$ fluxes turned
on%
\footnote{We also assume $h_{-}^{(1,1)}=0$, since we want to describe the simplest
situation.%
}. We refer to these vacua as {}``GKP compactifications.''

In these theories the vector moduli (descended from $\mathcal{N}=2$
vector multiplets) are the complex structure moduli $z^{i}$. The
hypermoduli (descended from $\mathcal{N}=2$ hypermultiplets) are
the axio-dilaton $\tau$ and the Kähler moduli $T_{\alpha}$.

At large volume and weak coupling, the Kähler potential factorizes
into 
\begin{equation}
K=K_{z}\left(z^{i}\right)-\log\left[-i\left(\tau-\overline{\tau}\right)\right]+K_{T}\left(T_{\alpha}\right),\label{eq:Simple-KP}
\end{equation}
 and enjoys a homogeneity property 
\begin{equation}
K_{\alpha\overline{\beta}}(\partial^{\alpha}K)(\partial^{\overline{\beta}}K)=3\,,\label{eq:Simple-Homogeneity}
\end{equation}
 where $K_{\alpha\overline{\beta}}$ is the $T_{\alpha}$ block of
the inverse Kähler metric. This homogeneity property is proven in
Appendix \ref{sec:Homogeneity}, as is a more general version. The
3-form RR and NS fluxes $F_{3}$ and $H_{3}$ give rise to the GVW
superpotential \citep{Gukov:1999ya,Taylor:1999ii} 
\begin{eqnarray}
W & = & \int G_{3}\wedge\Omega_{3}\,,
\end{eqnarray}
 where 
\begin{eqnarray}
G_{3} & \equiv & F_{3}-\tau H_{3}\,.
\end{eqnarray}
 The superpotential in this situation is linear in $\tau,$ and independent
of $T_{\alpha}.$

The scalar potential is 
\begin{eqnarray}
e^{-K}V & = & \sum_{X,Y=i,\tau,\alpha}K^{X\overline{Y}}D_{X}W\overline{D_{Y}W}-3\left|W\right|^{2},\label{eq:StandardPotential}
\end{eqnarray}
 with the Kähler derivative defined as 
\begin{equation}
D_{X}W=\partial_{X}W+W\partial_{X}K\,.
\end{equation}
 Because the superpotential is independent of the $T_{\alpha},$ and
because the Kähler potential \eqref{eq:Simple-KP} factorizes, the
4D scalar potential \eqref{eq:StandardPotential} reduces to 
\begin{eqnarray}
e^{-K}V & = & K^{i\overline{j}}D_{i}W\overline{D_{j}W}+K^{\tau\overline{\tau}}D_{\tau}W\overline{D_{\tau}W}+\left[K_{\alpha\overline{\beta}}\left(\partial^{\alpha}K\right)(\overline{\partial^{\beta}K})-3\right]\left|W\right|^{2}\\
 & = & K^{i\overline{j}}D_{i}W\overline{D_{j}W}+K^{\tau\overline{\tau}}D_{\tau}W\overline{D_{\tau}W}\,.\label{eq:Simple-No-Scale}
\end{eqnarray}
 The quantity in square brackets vanishes by virtue of the homogeneity
relation \eqref{eq:Simple-Homogeneity}. The inverse Kähler metric
$K^{i\overline{j}}$ has positive eigenvalues, and $K^{\tau\overline{\tau}}=4\mathrm{Im}\left(\tau\right)^{2}$
is positive, so this potential is positive semi-definite, with an
absolute minimum when $D_{i}W=0$ and $D_{\tau}W=0.$ Since $D^{\alpha}W=W\partial^{\alpha}K$
is generically non-zero, supersymmetry is broken. The combination
of supersymmetry breaking and vanishing scalar potential are the defining
features of a no-scale vacuum.

The linearity of the superpotential in $\tau$ allows us to write the
$D_{\tau}W=0$ condition in a more illuminating manner: 
\begin{eqnarray}
D_{\tau}W & = & -\int H_{3}\wedge\Omega_{3}-\frac{1}{\tau-\overline{\tau}}\int G_{3}\wedge\Omega_{3}\\
 & = & -\frac{1}{\tau-\overline{\tau}}\int\overline{G}_{3}\wedge\Omega_{3}=0\,.\label{eq:Simple-DtauW}
\end{eqnarray}
 If we combine this with 
\begin{equation}
D_{i}W=\int G_{3}\wedge D_{i}\Omega_{3}=0\,,\label{eq:Simple-DiW}
\end{equation}
 we see that the $\left(3,0\right)$ and $\left(1,2\right)$ pieces
of the complex flux $G_{3}$ must vanish. This is equivalent to the
condition that $G_{3}$ must be imaginary self-dual (ISD). In the
following we will find that analogues of \eqref{eq:Simple-DtauW}
and \eqref{eq:Simple-DiW}, and the resulting ISD conditions, arise
quite generically.

\subsection{\label{sub:GeneralCompactifications}General Type II, $\mathcal{N}=1$
Compactifications}

The GKP compactifications are very special, but the homogeneity and
linearity properties we used above apply to virtually all Type II
$\mathcal{N}=1$ flux compactifications, at least in the limit of
large volume and small coupling. In the general setting we will assume
that the moduli split into three groups: 
\begin{description}
\item [{Vector~moduli:}] these are descended from $\mathcal{N}=2$ vector
multiplets, and denoted $z^{i}.$ 
\item [{Stabilizable~hypermoduli:}] these appear linearly in the superpotential,
and are descended from $\mathcal{N}=2$ hypermultiplets. We denote
them by $t^{\hat{a}}.$ 
\item [{Unstabilizable~hypermoduli:}] these do not appear in the superpotential,
and are descended from $\mathcal{N}=2$ hypermultiplets. We denote
them by $t^{\hat{\alpha}}.$ 
\end{description}
We will denote all of the hypermoduli together as $t^{A}$, with the
$A$ index running over both ${\hat{a}}$ and ${\hat{\alpha}}$. The
split of the $t^{A}$ into $t^{\hat{a}}$ and $t^{\hat{\alpha}}$
will depend on which fluxes we have turned on. In the simple GKP example
we turned on $F_{3}$ and $H_{3},$ making the superpotential linear
in $\tau$. In this context $\tau$ is a stabilizable hypermodulus
while the $T_{\alpha}$, which did not appear in the superpotential, are
unstabilizable hypermoduli. Later, we will introduce $h_{-}^{(1,1)}$
moduli denoted $G^{a}$ which will appear linearly in the superpotential
due to a coupling to geometric fluxes. Then $\tau,G^{a}$ will all
be stabilizable hypermoduli in the terminology used here.

The most generic superpotential linear in the $t^{\hat{a}}$ and independent
of the $t^{\hat{\alpha}}$ can be written as: 
\begin{equation}
W=F\left(z\right)-\hat{t^{a}}H_{\hat{a}}\left(z\right).
\end{equation}
 Both $F\left(z\right)$ and $H_{\hat{a}}\left(z\right)$ are holomorphic
functions of the vector moduli $z^{i}$ and independent of the hypermoduli
$t^{A}.$ They can be thought of as generalizations of $\int F_{3}\wedge\Omega_{3}$
and $\int H_{3}\wedge\Omega_{3}.$

In the general setting we assume that the Kähler potential decomposes
into a term for the vector moduli $z^{i}$ and another term for all
the hypermoduli $t^{A}$, and enjoys the homogeneity relation: 
\begin{eqnarray}
K^{A\overline{B}}\left(\partial_{A}K\right)\left(\partial_{\overline{B}}K\right) & = & 4\,,\label{eq:homog1}\\
K^{A\overline{B}}\left(\partial_{\overline{B}}K\right) & = & \overline{t}^{A}-t^{A}\equiv-2i\eta^{A}\,.\label{eq:homog2}
\end{eqnarray}
 In the simple GKP example the Kähler potential \eqref{eq:Simple-KP}
had independent terms for $\tau$ and for the $T_{\alpha}$, but generally
it does not decompose so neatly. In that example $K^{\tau\overline{\tau}}\left(\partial_{\tau}K\right)\left(\partial_{\overline{\tau}}K\right)=1,$
so \eqref{eq:Simple-Homogeneity} is consistent with the homogeneity
\eqref{eq:homog1}. We discuss how these homogeneity relations arise in different
kinds of Type II compactifications in Appendix \ref{sec:Homogeneity}.

We include a brief aside on how corrections may affect our assumptions.
String theory corrections are generally governed by two expansion
parameters, the string coupling $g_{s}$, and the string tension $\al'$,
and quantities can receive corrections both perturbative and non-perturbative
in these parameters. For our models, the linearity of the superpotential
in the hypermoduli will hold perturbatively to all orders, but can
and will receive non-perturbative corrections (in both $g_{s}$ and
$\al'$) which we will neglect. The Kähler potential receives perturbative
corrections in both parameters which will generically ruin our homogeneity
and no-scale properties. However, if we stay at tree level in the
string coupling $g_{s}$, the $\al'$ corrections to the Kähler potential
still preserve the homogeneity property \eqref{eq:homog1}; for instance,
the first correction, calculated in~\citep{Becker:2002nn} simply
adds a term to $e^{-K}$ which is quartic in $\eta^{\tau}=\mathrm{Im}(\tau)$. The first
$g_{s}$ correction, however, not only ruins the homogeneity property,
but in fact mixes scalars coming from hypermultiplets with those coming
from vector multiplets. However, these $g_{s}$ corrections are also
typically accompanied by $\al'$ corrections (see for example~\citep{Berg:2007wt}).

The no-scale cancellation in the previous subsection involved the
$D^{\alpha}W$ terms in the scalar potential and the $-3\left|W\right|^{2}$
term. We are interested in similar cancellations in a more general
context, so we focus our attention on the $D_{A}W$ and $-3\left|W\right|^{2}$
terms. We can use the homogeneity relations \eqref{eq:homog1} and
\eqref{eq:homog2} to simplify them: 
\begin{eqnarray}
K^{A\overline{B}}D_{A}W\overline{D_{B}W}-3\left|W\right|^{2} & = & K^{\hat{a}\overline{\hat{b}}}D_{\hat{a}}W\overline{D_{\hat{b}}W}+K^{\hat{\alpha}\overline{\hat{\beta}}}\left(\partial_{\hat{\alpha}}K\right)\left(\partial_{\overline{\hat{\beta}}}K\right)\left|W\right|^{2}-3\left|W\right|^{2}\nonumber \\
 &  & +\left[K^{\hat{a}\overline{\hat{\alpha}}}\left(\partial_{\overline{\hat{\alpha}}}K\right)\overline{W}D_{\hat{a}}W+\mathrm{c.c.}\right]\,\\
 & = & K^{\hat{a}\overline{\hat{b}}}\left[D_{\hat{a}}W\overline{D_{\hat{b}}W}-\left(\partial_{\hat{a}}K\right)\left(\partial_{\overline{\hat{b}}}K\right)\left|W\right|^{2}\right]+\left|W\right|^{2}\nonumber \\
 &  & +\left[K^{\hat{a}\overline{\hat{\alpha}}}\left(\partial_{\overline{\hat{\alpha}}}K\right)\overline{W}\partial_{\hat{a}}W+\mathrm{c.c.}\right]\,\\
 & = & K^{\hat{a}\overline{\hat{b}}}\partial_{\hat{a}}W\overline{\partial_{\hat{b}}W}+\left|W\right|^{2}+\left[K^{\hat{a}\overline{A}}\left(\partial_{\overline{A}}K\right)\overline{W}\partial_{\hat{a}}W+\mathrm{c.c.}\right]\\
 & = & K^{\hat{a}\overline{\hat{b}}}\partial_{\hat{a}}W\overline{\partial_{\hat{b}}W}+\left|W\right|^{2}-2i\eta^{\hat{a}}\left[\overline{W}\partial_{\hat{a}}W-\mathrm{c.c.}\right]\,.\label{eq:GenNoScale-Intermediate}
\end{eqnarray}
 In the second step we used \eqref{eq:homog1}. 
 We then expanded out the $K^{\hat{a}\overline{\hat{b}}}D_{\hat{a}}W\overline{D_{\hat{b}}W}$
term and rearranged terms so that we could apply \eqref{eq:homog2}.
Recall that $\eta^{\hat{a}}$ is just the imaginary part of $t^{\hat{a}}.$

We can evaluate the derivatives in \eqref{eq:GenNoScale-Intermediate}
by virtue of the linearity of the superpotential, 
\begin{equation}
\partial_{\hat{a}}W=-H_{\hat{a}}\left(z\right)~.
\end{equation}
 The remaining terms in \eqref{eq:GenNoScale-Intermediate} simplify
when written in terms of 
\begin{equation}
\widetilde{W}\equiv F\left(z\right)-\overline{t}^{\hat{a}}H_{\hat{a}}\left(z\right)=W+2i\eta^{\hat{a}}H_{\hat{a}}\left(z\right)~,
\end{equation}
 the natural generalization of $\int\overline{G}_{3}\wedge\Omega_{3}$
from the previous example. Adding in the remaining terms in the scalar
potential, we now find 
\begin{equation}
e^{-K}V=K^{i\overline{j}}D_{i}W\overline{D_{j}W}+\left|\widetilde{W}\right|^{2}+\left[K^{\hat{a}\overline{\hat{b}}}-4\eta^{\hat{a}}\eta^{\hat{b}}\right]H_{\hat{a}}\left(z\right)\overline{H}_{\hat{b}}\left(z\right)\,.\label{eq:GenNoScale}
\end{equation}
 This is the natural generalization of the no-scale potential \eqref{eq:Simple-No-Scale}.

While the $D_{i}W$ and $\widetilde{W}$ terms in \eqref{eq:GenNoScale}
are closely related to the ISD conditions in the O3/O7 example, the
final set of terms is new. They will make a positive or negative contribution
to the potential depending on the eigenvalues of 
\begin{equation}
h^{\hat{a}\overline{\hat{b}}}\equiv K^{\hat{a}\overline{\hat{b}}}-4\eta^{\hat{a}}\eta^{\hat{b}}\,.\label{eq:hab-def}
\end{equation}
 The eigenvalues of $h^{\hat{a}\overline{\hat{b}}}$ are in general
functions of the hypermoduli. When $h^{\hat{a}\overline{\hat{b}}}$
has one or more negative eigenvalues, the scalar potential \eqref{eq:GenNoScale}
may admit AdS minima; we have little to say about such minima at this
time. However, when the eigenvalues of $h^{\hat{a}\overline{\hat{b}}}$
are positive semi-definite functions of the hypermoduli, Minkowski
vacua arise if 
\begin{eqnarray}
D_{i}W & = & 0\,,\label{eq:GenNoScale-1}\\
\widetilde{W} & = & 0\,,\label{eq:GenNoScale-2}\\
H_{\hat{\underline{a}}}\left(z\right) & = & 0\,,\label{eq:GenNoScale-3}
\end{eqnarray}
 where the $\hat{\underline{a}}$ index runs over the non-zero eigenvalues
of $h^{\hat{a}\overline{\hat{b}}}$ \emph{only}. Whenever $W$ is
non-vanishing supersymmetry is broken because 
\begin{equation}
D_{\hat{\underline{a}}}W=H_{\hat{\underline{a}}}\left(z\right)+\left(\partial_{\hat{\underline{a}}}K\right)W=\left(\partial_{\hat{\underline{a}}}K\right)W\neq0.
\end{equation}
 Thus our solutions are generally no-scale vacua. 

Before performing a detailed analysis of \eqref{eq:GenNoScale-1}-\eqref{eq:GenNoScale-3},
we can ask when they are likely to have solutions. Trouble can arise
if \eqref{eq:GenNoScale-1}-\eqref{eq:GenNoScale-3} together constitute
more equations than we have moduli. This occurs in two cases: 
\begin{itemize}
\item If $h^{\hat{a}\overline{\hat{b}}}$ has more positive eigenvalues
than there are vector moduli $z^{i},$ we will not in general be able
to solve the relevant $H_{\hat{\underline{a}}}\left(z\right)=0$ conditions.
This is because the $H_{\hat{a}}\left(z\right)$ are functions of
the $z^{i}$ \emph{only}, not of the hypermoduli. 
\item If $h^{\hat{a}\overline{\hat{b}}}$ has \emph{strictly} positive eigenvalues,
then the number of fields $z^{i}$ and $t^{\hat{\underline{a}}}$
is equal to the number of conditions in $D_{i}W=0$ and $H_{\hat{\underline{a}}}\left(z\right)=0.$
Because of the additional $\widetilde{W}=0$ condition, we do not
in general expect to be able to reach the Minkowski vacuum. Instead,
we expect the overall factor of $e^{K}$ in the potential to lead
to runaway vacua. 
\end{itemize}
We therefore expect Minkowski vacua to arise when $h^{\hat{a}\overline{\hat{b}}}$
has at least one zero eigenvalue, no negative eigenvalues, and the
number of positive eigenvalues is not greater than the number of $z^{i}.$
The previous O3/O7 example falls into this category, since $K^{\tau\overline{\tau}}=4\mbox{Im}\left(\tau\right)^{2},$
and thus the only eigenvalue of $h^{\hat{a}\overline{\hat{b}}}$ is
$h^{\tau\overline{\tau}}=0$. When $h^{\hat{a}\overline{\hat{b}}}$
is positive semi-definite but does not satisfy these properties, we
expect the overall factor of $e^{K}$ in the scalar potential to lead
to runaway vacua. 

In order to illustrate the utility of our simplified form for the
scalar potential \eqref{eq:GenNoScale}, we will present a new set
of Minkowski vacua in the next section. We will arrive at these by
adding geometric flux to the O3/O7 compactifications described at
the beginning of this section. The geometric flux will allow us to
stabilize additional hypermoduli, which cannot be stabilized with
3-form flux alone. It also appears to lead to an infinite series of
distinct vacua, as well as the ability to tune the string coupling
to be arbitrarily small. \textbf{ }We will present the full $h^{A\overline{B}}$
matrix for O3/O7 compactifications, and show that the conditions \eqref{eq:GenNoScale-1}-\eqref{eq:GenNoScale-3}
can easily be converted into flux attractor equations.

\section{\label{sec:GeomAttractorEqs}Attractor Equations and Geometric Flux}

The axio-dilaton $\tau$ is the only hypermodulus that enters the
perturbative type IIB superpotential in the presence of RR fluxes
and the 3-form NS flux $H_{3}$. There are several options for the
addition of extra ingredients that give rise to dependence on more
hypermoduli. Vacua with generalized NS fluxes are appealing because
T-duality establishes their existence in simple cases, while mirror
symmetry suggests their existence in more complicated cases. These
duality considerations also largely determine how these fluxes must
appear in the $\N=1$ superpotential. In this section we derive stabilization
conditions for the hypermoduli in this context, with emphasis on geometric
fluxes (sometimes called metric fluxes).

\subsection{Superpotential with Generalized Flux}

\label{sub:geomflux}

For a long time it has been known that in a background with $H$-flux
that lies parallel to a circle (i.e. if the circle isometry contracted
with $H$ is non-zero), a T-duality along the circle will generate
a new solution in which some components of $H$-flux have been exchanged
for some non-constant components of the metric \citep{Buscher:1987sk}.
The effect of these new metric components can be thought of as a twisting
of the circle over the rest of the geometry, encoded in the Cartan
equation 
\begin{equation}
de^{i}=f_{jk}^{i}e^{j}\wedge e^{k}~.
\end{equation}
The coefficients $f_{jk}^{i}$ serve as analogs of $H_{ijk}$, the
components of the original $H$-flux. Indeed, upon reduction to four
dimensions, these components appear as parameters of the low-energy
theory in much the same way as $H_{ijk}$ do \citep{Scherk:1979zr,Kaloper:1998kr,Kaloper:1999yr}.

If there are more circle isometries, one might be able to perform
a further T-duality, converting some of the $f_{jk}^{i}$ into new
objects $Q_{k}^{ij}$ known as non-geometric fluxes. In the presence
of non-geometric fluxes, the string background no longer has the structure
of a geometric manifold, but can still be understood as torus fibers
varying over a base, where the transition functions between patches
include string dualities \citep{Hellerman:2002ax,Dabholkar:2002sy}.
From a low-energy perspective, the non-geometric nature of the background
isn't relevant, and the components $Q_{k}^{ij}$ appear in a natural
way in the superpotential. In fact, from the low energy perspective,
one is also tempted to include objects $R^{ijk}$, which would correspond
to T-dualizing all three legs of some $H$-flux. From a ten-dimensional
perspective, it's not clear whether these latter fluxes can in fact
be constructed (indeed it is not clear whether all possible configurations
of the other geometric and non-geometric fluxes can be engineered),
but the manner in which they would appear in the effective theory
is essentially determined by symmetry considerations. For a more detailed
discussion see the review \citep{Wecht:2007wu} and references therein.

For the purposes of studying the superpotential and the tadpole constraints,
it will be useful to introduce a slightly different organizational
scheme for generalized fluxes. In order to present this scheme, we
must first give a basis for the cohomology of the underlying Calabi-Yau
orientifold where each element has definite parity under the orientifold
involution. For the remainder of this section we will specialize to
O3/O7 compactifications of type IIB string theory and take the basis
for even forms: 
\begin{itemize}
\item The constant function $1$ and the volume form $\varphi$, both even
under the orientifold involution. 
\item The 2-forms $\mu_{\al}$ and their dual 4-forms $\widetilde{\mu}^{\al}$.
All are even under the orientifold (so $\al=1,\ldots,h_{+}^{1,1}$). 
\item The 2-forms $\om_{a}$ and their dual 4-forms $\widetilde{\om}^{a}$.
All are odd under the orientifold (so $a=1,\ldots,h_{-}^{1,1}$). 
\end{itemize}
We will also introduce symplectic bases for the 3-forms where each
element has definite parity under the orientifold involution: 
\begin{itemize}
\item $(\mathcal{A}_{\hat{I}},\mathcal{B}^{\hat{I}})$ are even (so ${\hat{I}}=1,\ldots,h_{+}^{2,1}$). 
\item $(\alpha_{I},\beta^{I})$ are odd (so $I=0,\ldots,h_{-}^{2,1}$).
The extra index value is because the $(3,0)$ and $(0,3)$ forms are
odd. 
\end{itemize}
Now, in compactifications with $H$-flux, it is often very useful
to replace the local expressions $H_{ijk}$ for the components of
$H_{3}$ with a global expansion 
\begin{equation}
H_{3}=m_{h}^{I}\alpha_{I}-e_{I}^{h}\beta^{I}~,\label{eq:hthreeexp}
\end{equation}
 where $m_{h}^{I}$ and $e_{I}^{h}$ are the magnetic and electric
components of the 3-form flux. To obtain the analogous expansions
for the geometric and non-geometric fluxes, one should recast the $H$-flux
not just as a 3-form, but as a linear operator which maps $p$-forms
to $(p+3)$-forms (by wedging with $H_{3}$). The geometric fluxes
$f_{jk}^{i}$ similarly define a map from $p$-forms to $(p+1)$-forms,
while the non-geometric fluxes $Q_{k}^{ij}$ and $R^{ijk}$ give maps
from $p$-forms to $(p-1)$- and $(p-3)$-forms, respectively. Altogether,
we can combine these linear maps into an operator $\mathcal{D}$ \citep{Shelton:2006fd,Micu:2007rd},
which we can view as an operator of odd degree on the basis forms
of the underlying space. In particular we can write expansions of
$\mathcal{D}$ acting on the even forms 
\begin{eqnarray}
-\mathcal{D}\cdot1 & = & H_{3}=m_{h}^{I}\alpha_{I}-e_{I}^{h}\beta^{I}~,\non\label{eq:Donedef}\\
-\mathcal{D}\mu_{\al} & = & \widehat{r}_{\al}=\widehat{r}_{\al}^{\hat{I}}\mathcal{A}_{\hat{I}}-\widehat{r}_{\al\hat{I}}\mathcal{B}^{\hat{I}}~,\non\\
-\mathcal{D}\om_{a} & = & r_{a}=r_{a}^{I}\alpha_{I}-r_{aI}\beta^{I}~,\\
-\mathcal{D}\widetilde{\mu}^{\al} & = & \widehat{q}^{\al}=\widehat{q}^{\al I}\alpha_{I}-\widehat{q}_{I}^{\al}\beta^{I}~,\non\\
-\mathcal{D}\widetilde{\om}^{a} & = & q^{a}=q^{a{\hat{I}}}\mathcal{A}_{\hat{I}}-q_{\hat{I}}^{a}\mathcal{B}^{\hat{I}}~,\non\\
-\mathcal{D}\varphi & = & s=s^{\hat{I}}\mathcal{A}_{\hat{I}}-s_{\hat{I}}\mathcal{B}^{\hat{I}}~.\non\label{eq:Dphidef}
\end{eqnarray}
 The point here is just that $H_{3}$ and $Q_{k}^{ij}$ reverse the
parity of forms under the orientifold projection, while $f_{jk}^{i}$
and $R^{ijk}$ preserve it. We will not need the detailed map between
the component fluxes $f_{jk}^{i},Q_{k}^{ij},R^{ijk}$ and the 3-forms
$H_{3},\widehat{r}_{\al},r_{a},\widehat{q}^{\al},q^{a},s$ (given
in \citep{Ihl:2007ah}) because we will use only the latter terminology
from here on. For completeness, we note that there is of course also
an action analogous to \eqref{eq:Donedef}-\eqref{eq:Dphidef} on
the odd degree cohomology, but again we do not need the details.

Now, it turns out that the $H$-flux, the geometric fluxes labeled
$r_{a}$, and the non-geometric fluxes labeled $\widehat{q}^{\al}$
all contribute to the superpotential, while the geometric fluxes $\widehat{r}_{\al}$
and the non-geometric fluxes $q^{a}$ and $s$ contribute to D-terms~\citep{Robbins:2007yv}.
For the rest of this paper, we will focus on only the fluxes that
enter the superpotential, and set the latter group of fluxes to zero.

The operator $\mathcal{D}$ can be viewed as a generalization of the
twisted exterior derivative $d_{H}=d-H_{3}\w$~, which is the natural
differential operator on forms in the presence of $H_{3}$-flux. For
generalized fluxes, we would replace this with $\mathcal{D}$ and,
when acting on $d$-closed forms, we are left with just the linear
action \eqref{eq:Donedef}-\eqref{eq:Dphidef} of $\mathcal{D}$.
For consistency, the operator $\mathcal{D}$ must be nilpotent, $\mathcal{D}^{2}=0$,
like the usual exterior derivative~\citep{Shelton:2006fd,Ihl:2007ah}.
This constraint implies that the set of 3-forms $H_{3}$, $r_{a}$,
and $\widehat{q}^{\al}$ are all symplectically orthogonal%
\footnote{In the case at hand, where only $H_{3}$, $r_{a}$, $\widehat{q}^{\al}$
are nonzero, demanding that $\mathcal{D}^{2}=0$ on the cohomology
of the underlying space is equivalent to the condition that the 3-forms
form a symplectically orthogonal set. However, we actually need to
demand that $\mathcal{D}^{2}=0$ on locally defined closed forms,
and this requirement can be slightly more stringent. In this paper
we shall only make use of the symplectic orthogonality conditions,
with the understanding that our generalized fluxes may be somewhat
more constrained.%
}, i.e. that~\citep{Robbins:2007yv} 
\begin{equation}
\int H_{3}\w r_{a}=\int H_{3}\w\widehat{q}^{\al}=\int r_{a}\w r_{b}=\int r_{a}\w\widehat{q}^{\al}=\int\widehat{q}^{\al}\w\widehat{q}^{\beta}=0~,\label{eq:NS-Tadpole}
\end{equation}
 for all $a$, $b$, $\al$, $\beta$. Another perspective on these
constraints is to view them as NS source tadpole equations. For instance,
$\int H_{3}\w r_{a}$ contributes to the tadpole equation of NS5-branes
wrapping the two-cycle labeled by $a$, while $\int r_{a}\w r_{b}$
represents KK-monopole charge, , and other combinations correspond
to more exotic sources~\citep{Villadoro:2007tb}. Our models will
not include any of these NS sources, so the condition of symplectic
orthogonality stands.

Let us now briefly describe the hypermoduli, and the manner in which
they descend from $\N=2$ hypermultiplets \citep{Grimm:2004uq}. For
a type IIB O3/O7 compactification they are: 
\begin{itemize}
\item $\tau=C_{0}+ie^{-\phi}$, the axio-dilaton. 
\item $G^{a}$, $a=1,\ldots,h_{-}^{1,1}$. These arise from the complexified
2-form potential $C_{2}-\tau B_{2}=(c^{a}-\tau u^{a})\om_{a}=G^{a}\om_{a}$.
There is one of these for each 2-form $\om_{a}$ which is odd under
the orientifold involution. 
\item $T_{\al}$, $\al=1,\ldots,h_{+}^{1,1}$. These are obtained by expanding
a certain 4-form built out of the RR potential $C_{4}=\rho_{\al}\widetilde{\mu}^{\al}$
as well as the Kähler form $J=v^{\al}\mu_{\al}$. 
\end{itemize}
In fact, all of these hypermoduli can be conveniently and succinctly
obtained by expansion of a formal sum of even degree forms \citep{Benmachiche:2006df},

\begin{equation}
\Phi_{c}=e^{-B}\w C_{RR}+ie^{-\phi}\left(e^{-B+iJ}\right)=\tau+G^{a}\om_{a}+T_{\al}\widetilde{\mu}^{\al}~,\label{eq:IIBComplexSpinor}
\end{equation}
 where $C_{RR}=C_{0}+C_{2}+C_{4}$ is a formal sum of RR potentials.

We can now write down the perturbative superpotential in the presence
of the generalized fluxes. It takes exactly the same form as the familiar
GVW superpotential \citep{Gukov:1999ya}: 
\begin{equation}
W=\int G_{3}\w\Om_{3}~,\label{eq:superpot}
\end{equation}
 where 
\begin{equation}
\Om_{3}=Z^{I}\alpha_{I}-F_{I}\beta^{I}~,\label{eq:Omegaexpa}
\end{equation}
 is the usual holomorphic 3-form. We generalize from $G_{3}=F_{3}-\tau H_{3}$
in the GVW case to 
\begin{equation}
G_{3}=F_{3}+\mathcal{D}\Phi_{c}=F_{3}-\tau H_{3}-G^{a}r_{a}-T_{\al}\widehat{q}^{\al}~,\label{eq:gengthree}
\end{equation}
 when all the hypermoduli are taken into account. It is often useful
to present this complex flux in terms of components. Generalizing
\eqref{eq:hthreeexp} we expand the complex flux on basis 3-forms as 
\begin{equation}
G_{3}=m^{I}\alpha_{I}-e_{I}\beta^{I}~,
\end{equation}
 where now the complex flux components are 
\begin{eqnarray}
m^{I} & = & m_{f}^{I}-\tau m_{h}^{I}-G^{a}r_{a}^{I}-T_{\alpha}\hat{q}^{\al I}\,,\label{eq:mIdef}\\
e_{I} & = & e_{I}^{f}-\tau e_{I}^{h}-G^{a}r_{aI}-T_{\alpha}\hat{q}_{I}^{\al}\,.\label{eq:eIdef}
\end{eqnarray}
 The complex flux $G_{3}$ is a combination of the fluxes, $F_{3},H_{3},r_{a},\widehat{q}^{\al}$
that we consider {}``inputs,'' parts of the definition of the vacuum, and
then the hypermoduli $\tau,G^{a},T_{\alpha}$ which constitute dynamical
fields. 

The superpotential in component form is 
\begin{equation}
W=e_{I}Z^{I}-m^{I}F_{I}~.\label{eq:superpo}
\end{equation}
It is worth emphasizing that the superpotential depends on vector
moduli (complex structure moduli) and hypermoduli (Kähler moduli)
in quite different ways: 
\begin{itemize}
\item \textbf{Vector moduli:} enter through the symplectic section $(Z^{I},F_{I})$
in the familiar manner, described by special geometry and a holomorphic
prepotential $F$ with derivative $F_{I}$. The physical moduli can
(in one patch) be taken as the ratios $z^{i}=Z^{i}/Z^{0}$, $i=1,\ldots,h_{-}^{(2,1)}$. 
\item \textbf{Hypermoduli:} enter \textit{linearly} through the generalized
complex flux \eqref{eq:gengthree}. It is this property that we assumed
from the outset in the general discussion in section \ref{sub:GeneralCompactifications}. 
\end{itemize}

\subsection{\label{sub:hab}Spacetime Potential with Geometric Flux}

We next compute the spacetime potential \eqref{eq:GenNoScale} from
the superpotential \eqref{eq:superpo}. For this we need the Kähler
potential for the hypermoduli which, at large volume, is essentially
the volume of the compactification manifold 
\begin{equation}
K_{\mathrm{H}}=-\log[-i(\tau-\bar{\tau})^{4}\left(\mathcal{V}_{6}\right)^{2}]~.
\end{equation}
The Calabi-Yau volume $\mathcal{V}_{6}$ (equal to $(\kappa v^{3})/6$
in the notation below) depends implicitly on the hypermoduli $\tau,G^{a},T_{\alpha}$,
so it requires some effort to carry out differentiations with respect
to these scalar fields and obtain the Kähler metric. The final result
for the inverse Kähler metric becomes \citep{Grimm:2004uq} 
\begin{eqnarray}
K^{\tau\bar{\tau}} & = & -\left(\tau-\bar{\tau}\right)^{2}~,\non\label{eq:InverseMetricBegin}\\
K^{\tau\bar{a}} & = & \left(\tau-\bar{\tau}\right)^{2}u^{a}~,\non\\
K^{\tau}\vphantom{K}_{\bar{\al}} & = & -\hlf\left(\tau-\bar{\tau}\right)^{2}\left(\whk u^{2}\right)_{\al}~,\non\\
K^{a\bar{b}} & = & \left(\tau-\bar{\tau}\right)^{2}\left[\frac{1}{6}\left(\k v^{3}\right)\left(\whk v\right)^{-1\, ab}-u^{a}u^{b}\right]~,\\
K^{a}\vphantom{K}_{\bar{\al}} & = & \left(\tau-\bar{\tau}\right)^{2}\left[-\frac{1}{6}\left(\k v^{3}\right)\left[\left(\whk u\right)\left(\whk v\right)^{-1}\right]_{\al}^{a}+\hlf u^{a}\left(\whk u^{2}\right)_{\al}\right],\non\\
K_{\al\bar{\beta}} & = & \left(\tau-\bar{\tau}\right)^{2}\left[\frac{1}{6}\left(\k v^{3}\right)\left(\k v\right)_{\al\beta}-\frac{1}{4}\left(\k v^{2}\right)_{\al}\left(\k v^{2}\right)_{\beta}\right.\non\nonumber \\
 &  & \qquad\left.+\frac{1}{6}\left(\k v^{3}\right)\left[\left(\whk u\right)\left(\whk v\right)^{-1}\left(\whk u\right)\right]_{\al\beta}-\frac{1}{4}\left(\whk u^{2}\right)_{\al}\left(\whk u^{2}\right)_{\beta}\right].\non\label{eq:InverseMetricEnd}
\end{eqnarray}
 Here we have introduced intersection numbers 
\begin{equation}
\int\mu_{\al}\w\mu_{\beta}\w\mu_{\g}=\k_{\al\beta\g}~,\qquad\int\mu_{\al}\w\om_{a}\w\om_{b}=\whk_{\al\, ab}~,\label{eq:IntersectionNumbers}
\end{equation}
 and used a shorthand notation for contractions 
\begin{equation}
\left(\k v^{3}\right)=\k_{\al\beta\g}v^{\al}v^{\beta}v^{\g}~,\qquad\left(\whk v\right)_{ab}=\whk_{\al\, ab}v^{\al}~,\qquad\mathrm{etc.}\label{eq:IntersectionAbbreviations}
\end{equation}
 The spacetime potential \eqref{eq:GenNoScale} depends on the matrix
$h^{A{\bar{B}}}$ introduced in \eqref{eq:hab-def}, which is essentially
determined by the inverse Kähler metric. The fields $\eta^{A}$ are the imaginary
part of the hypermoduli, here 
\begin{eqnarray}
2i\eta^{\tau} & = & \tau-\bar{\tau}=2ie^{-\phi}\,,\\
2i\eta^{a} & = & -(\tau-\overline{\tau})u^{a}\,,\\
2i\eta_{\alpha} & = & \frac{\tau-\overline{\tau}}{2}\left[\left(\widehat{\kappa}u^{2}\right)_{\alpha}-\left(\kappa v^{2}\right)_{\alpha}\right]\,.
\end{eqnarray}
With this information, we easily find the matrix \eqref{eq:hab-def}:

\begin{eqnarray}
h^{\tau\bar{\tau}} & = & 0~,\non\\
h^{\tau\bar{a}} & = & 0~,\non\\
h_{\hphantom{\tau}\bar{\al}}^{\tau} & = & 2e^{-2\phi}\left(\k v^{2}\right)_{\al}~,\non\\
h^{a\bar{b}} & = & -\frac{2}{3}e^{-2\phi}\left(\k v^{3}\right)\left(\whk v\right)^{-1\, ab}~,\\
h_{\hphantom{a}\bar{\al}}^{a} & = & \frac{2}{3}e^{-2\phi}\left(\k v^{3}\right)\left[\left(\whk u\right)\left(\whk v\right)^{-1}\right]_{~\al}^{a}-2e^{-2\phi}u^{a}\left(\k v^{2}\right)_{\al}~,\non\\
h_{\al\bar{\beta}} & = & e^{-2\phi}\left\{ -\frac{2}{3}\left(\k v^{3}\right)\left(\k v\right)_{\al\beta}-\frac{2}{3}\left(\k v^{3}\right)\left[\left(\whk u\right)\left(\whk v\right)^{-1}\left(\whk u\right)\right]_{\al\beta}\right.\nonumber \\
 &  & \qquad\left.\vphantom{\frac{2}{3}}+\left(\k v^{2}\right)_{\al}\left(\whk u^{2}\right)_{\beta}+\left(\whk u^{2}\right)_{\al}\left(\k v^{2}\right)_{\beta}\right\} ~.\non
\end{eqnarray}
The vanishing of the components $h^{\tau\bar{\tau}}=h^{\tau\bar{a}}=0$
is significant. It means that, if we consider just the $\tau$ and
$G^{a}$ hypermoduli then $h^{\hat{a}\overline{\hat{b}}}$ has one zero eigenvalue.
Moreover, its remaining eigenvalues are positive, since $(\whk v)_{ab}$
is a negative-definite symmetric matrix inside the Kähler cone%
\footnote{This follows from the fact that the inverse Kähler metric above must
be positive definite at all points in the Kähler cone, and in particular
when $u^{a}=0$.%
}. According to the general criteria at the end of section \ref{sub:GeneralCompactifications}
this means that all of  $\tau$ and the $G^a$ would be stabilized. We are primarily interested in this setup, and will develop it further.

With the fluxes included in \eqref{eq:gengthree}, $T_{\alpha}$ is also
a stabilizable modulus. However, both $h_{\hphantom{a}\bar{\al}}^{a}$
and $h_{\alpha\overline{\beta}}$ have ambiguous signs, so including
all of the fluxes from \eqref{eq:gengthree} will generically lead
to AdS vacua. Since we are well-equipped to study Minkowski vacua,
we will set $\widehat{q}^{\alpha}=0$ for the remainder of the paper.
This reduces the complex flux $G_{3}$ from \eqref{eq:gengthree}
to 
\begin{eqnarray}
G_{3} & = & F_{3}-\tau H_{3}-G^{a}r_{a}\,,\label{eq:G3-noQ}
\end{eqnarray}
reduces the components of $G_{3}$ from \eqref{eq:mIdef} and \eqref{eq:eIdef}
to 
\begin{eqnarray}
m^{I} & = & m_{f}^{I}-\tau m_{h}^{I}-G^{a}r_{a}^{I}\,,\label{eq:mI-noQ}\\
e_{I} & = & e_{I}^{f}-\tau e_{I}^{h}-G^{a}r_{aI}\,,\label{eq:eI-noQ}
\end{eqnarray}
and renders $T_{\alpha}$ unstabilizable. 

For specific orientifold examples there can be other suitable truncations
which can include some of the $T_{\al}$. For instance, in a background
with $h_{-}^{1,1}=0$ and some particular even 2-form $\mu_{1}$ satisfying
$\mu_{1}\w\mu_{1}=0$ (for instance one can construct suitable examples
as certain complete intersections in products of projective spaces),
then $h_{-}^{\left(1,1\right)}=0$ and so we could truncate to $T_{1}$
alone (no $\tau$). However, such solutions are not generic.

\subsection{Attractor Equations from ISD Conditions}

We are now ready to derive the attractor equations that describe moduli
stabilization of O3/O7 compactifications with geometric flux as well
as conventional 3-form fluxes.

The starting point is a subset \eqref{eq:GenNoScale-1}-\eqref{eq:GenNoScale-2}
of the conditions for Minkowski vacua 
\begin{eqnarray}
D_{i}W & = & \int G_{3}\wedge D_{i}\Omega_{3}=0\,,\label{eq:ISD1}\\
\widetilde{W} & = & \int\overline{G}_{3}\wedge\Omega_{3}=0\,.\label{eq:ISD2}
\end{eqnarray}
The geometric flux $r_{a}$ and the hypermoduli $G^{a}$ just enter
through the complex flux $G_{3}$ \eqref{eq:G3-noQ}. The form of
the conditions \eqref{eq:ISD1}-\eqref{eq:ISD2} is therefore the
same as when there is no geometric flux. Indeed, these equations agree
with the ISD conditions \eqref{eq:Simple-DtauW} and \eqref{eq:Simple-DiW}
for O3/O7 compactifications with 3-form flux alone. As we will make
explicit, this means we can proceed as if there were no geometric
fluxes, and then determine the hypermoduli from the constraints \eqref{eq:GenNoScale-2}
and \eqref{eq:GenNoScale-3} at the end.

In the absence of geometric fluxes, it is known that \eqref{eq:ISD1}
and \eqref{eq:ISD2} are best analyzed in the complex basis $\left\{ \Omega_{3},D_{i}\Omega_{3},\overline{D_{i}\Omega}_{3},\overline{\Omega}_{3}\right\} $
for the 3-form cohomology. Symplectic orthogonality then determines
the complex flux $G_{3}$ as 
\begin{equation}
G_{3}=\overline{C\Omega}_{3}+C^{i}D_{i}\Omega_{3}\,,\label{eq:Gthreeexp}
\end{equation}
 with equality in the sense of cohomology. Since the complex basis
consists of eigenforms of the Hodge star ($*=+i$ on $\overline{\Omega}_{3}$,
$D_{i}\Omega_{3}$ and $*=-i$ on $\Omega_{3}$, $\overline{D_{j}\Omega}_{3}),$
it is manifest that $G_{3}$ is a generic ISD flux. The expansion
coefficients%
\footnote{The normalization of these is changed compared with \citep{Larsen:2009fw}:
$C_{\mathrm{here}}=i\mathrm{Im}\left(\tau\right)C_{\mathrm{there}},$
$C_{\mathrm{here}}^{i}=-i\mathrm{Im}\left(\tau\right)C_{\mathrm{there}}^{i},$
and $L_{\mathrm{here}}^{I}=-i\mathrm{Im}\left(\tau\right)L_{\mathrm{there}}^{I}.$ }%
 $C$ and $C^{i}$ determine the mass matrix for the moduli
\citep{Larsen:2009fw}.

Fluxes can be interpreted as twisting of the exterior derivative $d\to\mathcal{D}$,
as we have reviewed in section \ref{sub:geomflux}. The complex basis
$\left\{ \Omega_{3},D_{i}\Omega_{3},\overline{D_{i}\Omega}_{3},\overline{\Omega}_{3}\right\} $
is certainly a good basis for the 3-form cohomology of the underlying
Calabi-Yau \citep{Candelas:1990pi}, but relatively little is known
about the corresponding twisted cohomology. We can justify the continued
use of the complex basis by observing that the fluxes we consider
preserve $SU(3)$ structure, even though they generally spoil the
$SU(3)$ holonomy. The basis elements $\Omega_{3}$, $D_{i}\Omega_{3}$
transform in representations of $SU(3)$; the $SU\left(3\right)$
structure ensures that they satisfy the usual orthogonality relations,
and that they retain their eigenvalues under the Hodge star \citep{Grana:2006kf,Cassani:2007pq}.
We can therefore apply \eqref{eq:Gthreeexp} also after the introduction
of geometric fluxes, with the equality holding up to terms that vanish
in the integral. 

The covariant derivative with respect to the $z^{i}$ that appears
in \eqref{eq:Gthreeexp} is awkward (because it obscures symplectic
invariance) and also presents challenges in practical computations
(because the Kähler potential enters). It is advantageous to replace
it with an ordinary derivative with respect to the $Z^{I},$ i.e.

\begin{equation}
G_{3}=\overline{C\Omega}_{3}+L^{I}\partial_{I}\Omega_{3}\,.\label{eq:G3-2}
\end{equation}
In doing so we must be conscious of the fact that ordinary derivatives
of $\Omega_{3}$ contain a term proportional to $\Omega_{3}$: 
\begin{equation}
\partial_{I}\Omega_{3}=\left(\partial_{I}K\right)\Omega_{3}+\ldots\,.
\end{equation}
The $G_{3}$ \eqref{eq:Gthreeexp} cannot contain a term proportional
to $\Omega_{3}$ so we must impose an additional constraint: 
\begin{equation}
L^{I}\partial_{I}K=0\,,\label{eq:tau-constraint}
\end{equation}
on the $L^{I}$. There is indeed one more complex parameter among
the $L^{I}$ than there is among the $C^{i},$ which is consistent
with the addition of one complex constraint. Our result \eqref{eq:G3-2}
is the attractor equation, written as a relation between $3$-forms.

The attractor equations are perhaps more transparent when written
in terms of the real basis $(\alpha_{I},\beta^{I})$ of odd $3$-forms
introduced in section \ref{sub:geomflux}. Then the moduli are encoded
in the symplectic section $(Z^{I},F_{I})$ introduced in \eqref{eq:Omegaexpa}
and the flux components take the form \eqref{eq:mIdef}-\eqref{eq:eIdef}.
The component form of the attractor equation \eqref{eq:G3-2} becomes:

\begin{eqnarray}
m^{I} & = & \overline{CZ}^{I}+L^{I}\,,\label{eq:m-1}\\
e_{I} & = & \overline{CF}_{I}+L^{J}F_{IJ}\,.\label{eq:e-1}
\end{eqnarray}
We consider $CZ^{I}$ and $L^{I}$ to be the independent variables
in the attractor equations. The $CZ^{I}$ determine the physical moduli
$z^{i}$ as well as an additional parameter, $CZ^{0},$ which only
appears in the scalar mass matrix. The $L^{I}$ are all mass parameters.
$CF_{I}$ and $F_{IJ}$ are functions of the $CZ^{I}$ -- the specific
functional forms are determined by the symplectic section of the Calabi-Yau.
Since the number of attractor equations in \eqref{eq:m-1},\eqref{eq:e-1}
is equal to the number of variables in $CZ^{I}$, $L^{I},$ solving
\eqref{eq:m-1} and \eqref{eq:e-1} should give $CZ^{I}$ and $L^{I}$
as functions of the complex fluxes $m^{I}$ and $e_{I}.$ This is
true whether or not there are any geometric fluxes.

Now, this type of solution does not yet determine $CZ^{I}$ and $L^{I}$
as functions of the real, physical fluxes, because the complex fluxes
$(m^{I},e_{I})$ are themselves functions of the hypermoduli. This
dependence on the hypermoduli is governed by several constraints.
There is both the universal constraint \eqref{eq:tau-constraint},
written in components as 
\begin{equation}
0=\overline{CF}_{I}L^{I}-\overline{CZ}^{I}L^{J}F_{IJ}\,,\label{eq:tau-1}
\end{equation}
 and generally also the constraints \eqref{eq:GenNoScale-3}. When only geometric fluxes have been included, these latter constraints
are the conditions: 
\begin{equation}
H_{a}\left(z\right)=\int\Omega_{3}\wedge r_{a}=0\,,
\end{equation}
 which we can write in terms of components as 
\begin{equation}
0=r_{Ia}CZ^{I}-r_{a}^{I}CF_{I}\,.\label{eq:Ga-1}
\end{equation}
We emphasize that we have not set $D_{a}W=0,$ but instead that $D_{a}W=W\partial_{a}K$
leads to supersymmetry breaking when $W\neq0$. This stands in contrast
to the flux attractor equations for $SU\left(3\right)\times SU\left(3\right)$
structure compactifications developed in \citep{Anguelova:2008fm,Anguelova:2009az}.
These attractor equations only described supersymmetric $\left(W=0\right)$
Minkowski vacua, while the attractor equations presented here describe
non-supersymmetric $\left(W\neq0\right)$ Minkowski vacua as well.

Let us summarize the procedure we propose. We first solve the attractor
equations \eqref{eq:m-1} and \eqref{eq:e-1} for $CZ^{I}$, $L^{I}$.
The result will be in terms of the complex fluxes $(m^{I},e_{I})$
that depend on both $\tau$ and $G^{a}$. In the next step we use
the constraints \eqref{eq:tau-1} and \eqref{eq:Ga-1} together to
determine $\tau$ and the $G^{a}$. The procedure is particularly
simple in the standard GKP case where there is no geometric flux,
and so the complex fluxes depend only on $\tau$. Then there is just
a single constraint \eqref{eq:tau-1} to solve. In the remainder of
the paper we will study the more general case including geometric
fluxes.

There is one subtlety: although the constraints \eqref{eq:tau-1},
\eqref{eq:Ga-1} appear to determine all of $\tau$, $G^{a}$, in
fact the number of $\tau$, $G^{a}$ that we can stabilize is limited
by the number $h_{-}^{\left(2,1\right)}$ of physical moduli $z^{i}=Z^{i}/Z^{0}.$
If we divide \eqref{eq:Ga-1} by $CZ^{0}$ and use the homogeneity
properties of the $F_{I},$ we see that the hypermoduli enter into
\eqref{eq:Ga-1} only via the $z^{i},$ so only $h_{-}^{\left(2,1\right)}$
distinct combinations of the hypermoduli are constrained. When $h_{-}^{\left(1,1\right)}>h_{-}^{\left(2,1\right)},$
either $h_{-}^{\left(1,1\right)}-h_{-}^{\left(2,1\right)}$ hypermoduli
will remain unstabilized, or there will be no solutions to \eqref{eq:Ga-1}
and we are forced into a runaway vacuum.

The situation is ameliorated somewhat by that fact that not all of
the constraints \eqref{eq:Ga-1} can be independent. The geometric
fluxes $r_{a}$ are 3-forms that must be symplectically orthogonal
due to the tadpole conditions \eqref{eq:NS-Tadpole}. There are at
most $h_{-}^{\left(2,1\right)}+1$ such three-forms, so only $h_{-}^{\left(2,1\right)}+1$
of the constraints \eqref{eq:Ga-1} can be independent. This is still
one more than $h_{-}^{\left(2,1\right)}$, the number of $z^{i}$'s
and thus the number of independent equations we can solve, according
to the argument in the previous paragraph. For generic geometric fluxes
and $h_{-}^{\left(1,1\right)}>h_{-}^{\left(2,1\right)}$ we will therefore
find no solutions to \eqref{eq:Ga-1}, but for a codimension one
subspace of the space of possible geometric fluxes, we expect to be
effective at stabilizing hypermoduli.

\textbf{Summary of this section:} The principal results are the attractor
equations, \eqref{eq:m-1} and \eqref{eq:e-1}, and the constraints
\eqref{eq:tau-1} and \eqref{eq:Ga-1}. These equations illuminate
how particular fluxes stabilize particular moduli. In the following
sections we will show that solutions to these attractor equations
can be succinctly summarized by a single generating function, as was
the case without geometric fluxes. We will also solve several examples
where as many moduli as possible are stabilized.

\section{\label{sec:GenFnsGeom}Generating Functions with Geometric Flux}

While the flux attractor equations \eqref{eq:m-1}, \eqref{eq:e-1}
and constraints \eqref{eq:tau-1}, \eqref{eq:Ga-1}, are considerably
simpler than the equations that would arise from direct minimization
of the potential, they cannot be solved explicitly for a generic Calabi-Yau.
Nevertheless, we can establish several general properties of the solutions.
First of all, the solutions for all of the moduli and mass parameters
can be presented as derivatives of a single generating function. This
was first shown in \citep{Larsen:2009fw} for the standard GKP setup,
and here we extend the result to include geometric fluxes. 

We will present two versions of the generating function, which give
rise to two different stabilization procedures. The first version
depends on both the complex fluxes and the hypermoduli, with the stabilized
values of the hypermoduli determined by extremizing the generating
function with respect to the hypermoduli. The second version employs
a reduced generating function that depends on the real fluxes only.
In both cases the stabilization of the vector moduli is treated separately
from the stabilization of the hypermoduli.

\subsection{Explicit Expression for the Generating Function}

We begin by rewriting the electric and magnetic attractor equations,
\eqref{eq:m-1} and \eqref{eq:e-1} as: 
\begin{eqnarray}
\overline{CZ}^{I} & = & \frac{1}{2}\left(m^{I}+\phi^{I}\right),\label{eq:NewMag-1}\\
L^{I} & = & \frac{1}{2}\left(m^{I}-\phi^{I}\right),\label{eq:NewMag-2}\\
\overline{CF}_{I} & = & \frac{1}{2}\left(e_{I}+\theta_{I}\right),\label{eq:NewElec-1}\\
L^{J}F_{IJ} & = & \frac{1}{2}\left(e_{I}-\theta_{I}\right),\label{eq:NewElec-2}
\end{eqnarray}
where the $\phi^{I}$ and $\theta_{I}$ are (typically non-holomorphic)
functions of the complex fluxes $m^{I}$ and $e_{I}.$ Although it
may appear that arbitrary $\phi^{I}$ and $\theta_{I}$ solve \eqref{eq:m-1}
and \eqref{eq:e-1}, leading to essentially arbitrary solutions for
$CZ^{I}$ and $L^{I},$ the solutions for $\phi^{I}$ and $\theta_{I}$
are in fact related to one another in a nonlinear fashion. This is
because $F_{I}$ and $F_{IJ}$ are not independent parameters, but
are fixed functions of the $Z^{I},$ with the specific functional
form determined by the symplectic section of the Calabi-Yau. In order
to solve \eqref{eq:NewMag-1}-\eqref{eq:NewElec-2}, we must substitute
the expressions for $CZ^{I}$ and $L^{I}$ in terms of $m^{I}$ and
$\phi^{I}$ into \eqref{eq:NewElec-1} and \eqref{eq:NewElec-2},
then solve for $\phi^{I}$ and $\theta_{I}.$ Doing this directly
is difficult even for relatively simple Calabi-Yaus.

Considered as equations that determine the potentials $\phi^{I}$
and $\theta_{I}$ in terms of the complex fluxes $m^{I}$ and $e_{I},$
\eqref{eq:NewMag-1}-\eqref{eq:NewElec-2} are exactly the same whether
or not we have introduced geometric fluxes. We can therefore use a
result proven in \citep{Larsen:2009fw}, namely that all solutions
for the $\phi^{I}$ and $\theta_{I}$ can be written as derivatives
of a real generating function%
\footnote{Strictly speaking, we can arrive at a whole family of generating functions
by changing the normalizing factor of $\left(\tau-\overline{\tau}\right)$
to various other functions of the hypermoduli. In section \eqref{sub:HyperDerivatives}
we will see that the choice of $\left(\tau-\overline{\tau}\right)$
is preferred, even after we have introduced geometric fluxes.%
} $\mathcal{G}:$ 
\begin{eqnarray}
\phi^{I} & = & \left(\tau-\overline{\tau}\right)\frac{\partial\mathcal{G}}{\partial\overline{e}_{I}}\,,\label{eq:phi-ansatz}\\
\theta_{I} & = & -\left(\tau-\overline{\tau}\right)\frac{\partial\mathcal{G}}{\partial\overline{m}^{I}}\,.\label{eq:theta-ansatz}
\end{eqnarray}
Although the additional minus sign in \eqref{eq:theta-ansatz} may
look awkward, it is necessary because $\left(\partial/\partial\overline{e}_{I},-\partial/\partial\overline{m}^{I}\right)$
is a good symplectic vector, while $\left(\partial/\partial\overline{e}_{I},\partial/\partial\overline{m}^{I}\right)$
is not. The derivatives of $\mathcal{G}$ are taken with the other
\emph{complex} fluxes, as well as $\tau$ and the $G^{a},$ held fixed.
If we consider $\mathcal{G}$ as a thermodynamic function, \eqref{eq:phi-ansatz}
and \eqref{eq:theta-ansatz} identify $\phi^{I}$ and $\theta_{I}$
as the potentials conjugate to $\overline{e}_{I}$ and $\overline{m}^{I},$
respectively, and so we will frequently refer to them as {}``the
potentials.''

Another result of \citep{Larsen:2009fw} that still holds after the
introduction of geometric flux is that $\mathcal{G}$ is homogeneous
of degree $\left(1,1\right)$ in the complex fluxes. In other words,
\begin{eqnarray}
\mathcal{G}\left(\lambda m^{J},\lambda e_{J},\tilde{\lambda}\overline{m}^{J},\tilde{\lambda}\overline{e}_{J},\tau,G^{a}\right) & = & \lambda\tilde{\lambda}\mathcal{G}\left(m^{J},e_{J},\overline{m}^{J},\overline{e}_{J},\tau,G^{a}\right),
\end{eqnarray}
for any $\lambda,\tilde{\lambda}\in\mathbb{C}.$ This implies that
the potentials $\phi^{I}$ and $\theta_{I}$ are homogeneous of degree
$\left(1,0\right).$ It also allows us to write an explicit expression
for $\mathcal{G}:$
\begin{eqnarray}
\mathcal{G} & = & e_{I}\frac{\partial\mathcal{G}}{\partial e_{I}}+m^{I}\frac{\partial\mathcal{G}}{\partial m^{I}}\\
 & = & -\frac{1}{\tau-\overline{\tau}}\left\{ e_{I}\overline{\phi}^{I}-m^{I}\overline{\theta}_{I}\right\} \,.\label{eq:Gen-1b}
\end{eqnarray}
The first line follows from the homogeneity of $\mathcal{G},$ while
the second follows by substituting in \eqref{eq:phi-ansatz} and \eqref{eq:theta-ansatz}.
Given an explicit solution of \eqref{eq:NewMag-1}-\eqref{eq:NewElec-2},
we can compute $\phi^{I}$ and $\theta_{I},$ then use \eqref{eq:Gen-1b}
to compute $\mathcal{G}.$ We also see that whenever the flux attractor
equations have multiple sets of solutions, each solution will correspond
to a different generating function.

\subsection{\label{sub:HyperDerivatives}Stabilizing the Hypermoduli}

Once the potentials have been determined, we have solved the attractor
equations \eqref{eq:m-1} and \eqref{eq:e-1} for the unknowns $CZ^{I}$
and $L^{I},$ with the hypermoduli treated as given parameters. To
find the stabilized values of the hypermoduli we can substitute our
solutions for $CZ^{I}$ and $L^{I}$ into the constraints \eqref{eq:tau-1}
and \eqref{eq:Ga-1} and solve. In this section we will present an
alternate procedure: simply extremize $\mathcal{G}$ with respect
to the hypermoduli. 

We first present the universal constraint in a simplified form. If
we substitute \eqref{eq:NewMag-1}-\eqref{eq:NewElec-2} into \eqref{eq:tau-1},
we find
\begin{equation}
0=\overline{CF}_{I}L^{I}-\overline{CZ}^{I}L^{J}F_{IJ}=-\frac{1}{2}\left(\phi^{I}e_{I}-\theta_{I}m^{I}\right).\label{eq:tau-2}
\end{equation}
In order to recover the universal constraint and the constraints \eqref{eq:Ga-1}
from our new procedure, we need the derivatives of $\mathcal{G}$
with the real fluxes, rather than the complex fluxes, held fixed.

We begin by writing the $\tau-$derivative of $\mathcal{G}$ with
the real fluxes held fixed%
\footnote{The specific form of $\left.\frac{\partial\mathcal{G}}{\partial\tau}\right|_{\mathbb{C}}$
was ultimately determined by the introduction of $\left(\tau-\overline{\tau}\right),$
rather than some other function of the hypermoduli, in \eqref{eq:phi-ansatz}
and \eqref{eq:theta-ansatz}. Using $\left(\tau-\overline{\tau}\right)$
we will find simple conditions on $\left.\frac{\partial\mathcal{G}}{\partial\tau}\right|_{\mathbb{R}}$
and $\left.\frac{\partial\mathcal{G}}{\partial G^{a}}\right|_{\mathbb{R}},$
while using other functions of the hypermoduli would lead to much
more awkward conditions.%
} : 
\begin{eqnarray}
\left.\frac{\partial\mathcal{G}}{\partial\tau}\right|_{\mathbb{R}} & = & \left.\frac{\partial\mathcal{G}}{\partial\tau}\right|_{\mathbb{C}}+\frac{\partial\mathcal{G}}{\partial e_{I}}\frac{\partial e_{I}}{\partial\tau}+\frac{\partial\mathcal{G}}{\partial m^{I}}\frac{\partial m^{I}}{\partial\tau}\\
 & = & \frac{1}{\left(\tau-\overline{\tau}\right)^{2}}\left\{ \overline{\phi}^{I}e_{I}-\overline{\theta}_{I}m^{I}\right\} +\frac{1}{\tau-\overline{\tau}}\left\{ \overline{\phi}^{I}e_{I}^{h}-\overline{\theta}_{I}m_{h}^{I}\right\} \,.\label{eq:G-tau-1}
\end{eqnarray}
In the first line we used $\mathbb{R}$ and $\mathbb{C}$ as a shorthand
to indicate that the real fluxes and complex fluxes, respectively,
are held fixed. The second line follows by application of \eqref{eq:Gen-1b},
\eqref{eq:phi-ansatz}-\eqref{eq:theta-ansatz}, \eqref{eq:mI-noQ},
and \eqref{eq:eI-noQ}. In the standard GKP setup, this expression
reduces to
\begin{eqnarray}
\left.\frac{\partial\mathcal{G}}{\partial\tau}\right|_{\mathbb{R}} & = & \frac{1}{\left(\tau-\overline{\tau}\right)^{2}}\left\{ \overline{\phi}^{I}\left[e_{I}^{f}-\tau e_{I}^{h}+\left(\tau-\overline{\tau}\right)e_{I}^{h}\right]-\overline{\theta}^{I}\left[m_{f}^{I}-\tau m_{h}^{I}+\left(\tau-\overline{\tau}\right)m_{h}^{I}\right]\right\} \\
 & = & \frac{1}{\left(\tau-\overline{\tau}\right)^{2}}\left\{ \overline{\phi}^{I}\overline{e}_{I}-\overline{\theta}_{I}\overline{m}^{I}\right\} \,.\label{eq:dtau-R}
\end{eqnarray}
Comparing \eqref{eq:dtau-R} with \eqref{eq:tau-2}, we see that extremizing
$\mathcal{G}$ with respect to $\tau,$ while holding the real fluxes
fixed, reproduces \eqref{eq:tau-1} in the standard GKP setup.

After adding geometric fluxes, \eqref{eq:G-tau-1} reduces to
\begin{equation}
\left.\frac{\partial\mathcal{G}}{\partial\tau}\right|_{\mathbb{R}}=\frac{1}{\left(\tau-\overline{\tau}\right)^{2}}\left\{ \overline{\phi}^{I}\left[e_{I}^{f}-\overline{\tau}e_{I}^{h}-G^{a}r_{Ia}\right]-\overline{\theta}^{I}\left[m_{f}^{I}-\overline{\tau}m_{h}^{I}-G^{a}r_{a}^{I}\right]\right\} ,\label{eq:G-tau-2}
\end{equation}
so a $\tau-$derivative alone is insufficient to reproduce \eqref{eq:tau-2}.
However, we can combine \eqref{eq:G-tau-2} with
\begin{eqnarray}
\left.\frac{\partial\mathcal{G}}{\partial G^{a}}\right|_{\mathbb{R}} & = & \left.\frac{\partial\mathcal{G}}{\partial G^{a}}\right|_{\mathbb{C}}+\frac{\partial\mathcal{G}}{\partial e_{I}}\frac{\partial e_{I}}{\partial G^{a}}+\frac{\partial\mathcal{G}}{\partial m^{I}}\frac{\partial m^{I}}{\partial G^{a}}\\
 & = & \frac{1}{\tau-\overline{\tau}}\left\{ \overline{\phi}^{I}r_{Ia}-\overline{\theta}_{I}r_{a}^{I}\right\} \,,\label{eq:G-Ga-1}
\end{eqnarray}
to find
\begin{equation}
\left(\tau-\overline{\tau}\right)\left.\frac{\partial\mathcal{G}}{\partial\tau}\right|_{\mathbb{R}}+\left(G^{a}-\overline{G}^{a}\right)\left.\frac{\partial\mathcal{G}}{\partial G^{a}}\right|_{\mathbb{R}}=\frac{1}{\tau-\overline{\tau}}\left\{ \overline{\phi}^{I}\overline{e}_{I}-\overline{\theta}_{I}\overline{m}^{I}\right\} \,.\label{eq:G-tau-3}
\end{equation}
Setting this linear combination of derivatives of $\mathcal{G}$ to
zero thus reproduces \eqref{eq:tau-1}, even when geometric fluxes
are included. 

We also need to recover the remaining constraint \eqref{eq:Ga-1}
from derivatives of $\mathcal{G}.$ This is straightforward, because
the tadpole constraints \eqref{eq:NS-Tadpole} imply that 
\begin{equation}
\overline{m}^{I}r_{Ia}-\overline{e}_{I}r_{a}^{I}=0\,,
\end{equation}
and so allow us to rewrite \eqref{eq:G-Ga-1} as 
\begin{eqnarray}
\left.\frac{\partial\mathcal{G}}{\partial G^{a}}\right|_{\mathbb{R}} & = & \frac{1}{\tau-\overline{\tau}}\left\{ \left(\overline{m}^{I}+\overline{\phi}^{I}\right)r_{Ia}-\left(\overline{e}_{I}+\overline{\theta}_{I}\right)r_{a}^{I}\right\} \\
 & = & \frac{1}{\tau-\overline{\tau}}\left\{ CZ^{I}r_{Ia}-CF_{I}r_{a}^{I}\right\} \,.
\end{eqnarray}
Comparing this with \eqref{eq:Ga-1}, we see that extremizing $\mathcal{G}$
with respect to the $G^{a},$ while holding the real fluxes fixed,
reproduces the $H_{a}\left(z\right)=0$ attractor equations. Combining
this with \eqref{eq:G-tau-3}, we find that we must extremize over
$\tau$ as well. It is somewhat surprising that the tadpole constraints
play a crucial role here, given that they do not appear anywhere else
in our study of the flux attractor equations. 

Let us summarize our results about $\mathcal{G}$ so far. Suppose
that we have somehow determined $\mathcal{G}$ as a function of the
complex fluxes and the hypermoduli. \eqref{eq:phi-ansatz} and \eqref{eq:theta-ansatz}
then determine the potentials $\phi^{I}$ and $\theta_{I}$ as functions
of the complex fluxes and $\tau,$ and \eqref{eq:NewMag-1} and \eqref{eq:NewMag-2}
in turn determine the stabilized values of the vector moduli and mass
parameters. The remaining dependence of these quantities on the hypermoduli,
through the complex fluxes, is fixed by extremizing $\mathcal{G}$
with respect to the hypermoduli, while holding the real fluxes fixed.
Upon substituting the values of the hypermoduli into the expressions
for $\phi^{I}$ and $\theta_{I},$ we have determined the values of
all the moduli, as well as the values of the mass parameters $CZ^{0}$
and $L^{I}.$

\subsection{Reduced Generating Function}

One peculiar aspect of the generating function described so far is
that the fluxes and hypermoduli appear in $\mathcal{G}$ on roughly
equal footing, but are treated very differently when we solve for
the various moduli. We will now show how the moduli $z^{i}$ and mass
parameters $CZ^{0}$ and $L^{I}$ can be determined from a reduced
generating function, $\widetilde{\mathcal{G}},$ which depends on
the real fluxes only. Formally, $\widetilde{\mathcal{G}}$ is constructed
by substituting the stabilized values of the hypermoduli into $\mathcal{G}.$ 

We first address a preliminary issue concerning the map between real
and complex fluxes. While we have already recorded the expressions
for the complex fluxes in terms of the real fluxes \eqref{eq:mIdef}-\eqref{eq:eIdef},
we will also need to know how derivatives with respect to the complex
fluxes are related to derivatives with respect to the real fluxes,
and this relationship is slightly subtle. When discussing the real
fluxes we always explicitly include the full set $\left\{ m_{h}^{I},m_{f}^{I},e_{I}^{h},e_{I}^{f},r_{a}^{I},r_{aI}\right\} ,$
but when discussing the complex fluxes we tend to include only $m^{I},$
$e_{I},$ and their complex conjugates. In fact the complete set consists
of $\left\{ m^{I},\overline{m}^{I},e_{I},\overline{e}_{I},r_{a}^{I},r_{aI}\right\} .$
This implies that the relationship between the real and complex derivatives
is:
\begin{eqnarray}
\frac{\partial}{\partial m^{I}} & = & -\frac{1}{\tau-\overline{\tau}}\left(\overline{\tau}\frac{\partial}{\partial m_{f}^{I}}+\frac{\partial}{\partial m_{h}^{I}}\right),\label{eq:dm}\\
\frac{\partial}{\partial e_{I}} & = & -\frac{1}{\tau-\overline{\tau}}\left(\overline{\tau}\frac{\partial}{\partial e_{I}^{f}}+\frac{\partial}{\partial e_{I}^{h}}\right).\label{eq:de}
\end{eqnarray}
We might have expected derivatives with respect to $r_{a}^{I}$ or
$r_{aI}$ to appear here as well, but the complex derivatives must
give zero when acting on $r_{a}^{I}$ and $r_{aI},$ so such terms
cannot appear.

The decomposition of \eqref{eq:dm} and \eqref{eq:de} into derivatives
with respect to real fluxes suggests that we define a set of real
potentials,
\begin{eqnarray}
\phi_{f}^{I}=\frac{\partial\mathcal{G}}{\partial e_{I}^{h}}\,, &  & \phi_{h}^{I}=-\frac{\partial\mathcal{G}}{\partial e_{I}^{f}}\,,\label{eq:dG-phi}\\
\theta_{I}^{f}=-\frac{\partial\mathcal{G}}{\partial m_{h}^{I}}\,, &  & \theta_{I}^{h}=\frac{\partial\mathcal{G}}{\partial m_{f}^{I}}\,.\label{eq:dG-theta}
\end{eqnarray}
related to the complex potentials via
\begin{eqnarray}
\phi^{I} & = & \phi_{f}^{I}-\tau\phi_{h}^{I}\,,\label{eq:Real-Phi}\\
\theta_{I} & = & \theta_{I}^{f}-\tau\theta_{I}^{h}\,.\label{eq:Real-Theta}
\end{eqnarray}
Note that the derivatives with respect to the real fluxes are taken
with the hypermoduli held fixed. 

We now define $\widetilde{\mathcal{G}}$ as $\mathcal{G}$ with all
hypermoduli replaced by their stabilized values, written as functions
of the real fluxes. While this is the natural way to turn $\mathcal{G}$
into a function of the fluxes alone, we would like to know how $\widetilde{\mathcal{G}}$
relates to the attractor equations. A simple calculation shows that
\begin{eqnarray}
\frac{\partial\widetilde{\mathcal{G}}}{\partial e_{I}^{h}} & = & \frac{\partial\mathcal{G}}{\partial e_{I}^{h}}+\frac{\partial\mathcal{G}}{\partial\tau}\frac{\partial\tau}{\partial e_{I}^{h}}+\frac{\partial\mathcal{G}}{\partial G^{a}}\frac{\partial G^{a}}{\partial e_{I}^{h}}\\
 & = & \phi_{f}^{I}\,.
\end{eqnarray}
The second and third terms vanish because the hypermoduli are determined
by extremizing $\mathcal{G}$ with respect to $\tau$ and $G^{a}.$
We see that derivatives of $\widetilde{\mathcal{G}}$ return the real
potentials, and therefore determine the complex potentials $\phi^{I}$
and $\theta_{I}$ as functions of $\tau.$

The procedure we follow to determine the values of the moduli and
mass parameters if we know the reduced generating function $\widetilde{\mathcal{G}}$
is slightly different from the procedure we follow if we have $\mathcal{G}.$
We first differentiate $\widetilde{\mathcal{G}}$ to determine the
real potentials. This gives us the moduli and mass parameters as functions
of the real fluxes and the hypermoduli:
\begin{eqnarray}
\overline{CZ}^{I} & = & \frac{1}{2}\left[\left(m_{f}^{I}-\tau m_{h}^{I}-G^{a}r_{a}^{I}\right)+\left(\frac{\partial\widetilde{\mathcal{G}}}{\partial e_{I}^{h}}+\tau\frac{\partial\widetilde{\mathcal{G}}}{\partial e_{I}^{f}}\right)\right],\\
L^{I} & = & \frac{1}{2}\left[\left(m_{f}^{I}-\tau m_{h}^{I}-G^{a}r_{a}^{I}\right)-\left(\frac{\partial\widetilde{\mathcal{G}}}{\partial e_{I}^{h}}+\tau\frac{\partial\widetilde{\mathcal{G}}}{\partial e_{I}^{f}}\right)\right].
\end{eqnarray}
We then substitute these expressions into \eqref{eq:tau-1} and \eqref{eq:Ga-1}
and solve to find $\tau$ and the $G^{a}.$ 

We believe that $\widetilde{\mathcal{G}}$ is a conceptually simpler
object to study than $\mathcal{G}$ since it is a function of the
fluxes alone, rather than a function of fluxes and hypermoduli. We
will also see an example below where we cannot determine a closed form
for $\mathcal{G},$ but are able to compute $\widetilde{\mathcal{G}}.$

\section{Examples}

In order to establish the utility of the flux attractor equations
and the generating function formalism, we will now analyze two compactifications
that admit both 3-form fluxes and geometric fluxes. We will solve
the attractor equations \eqref{eq:m-1} and \eqref{eq:e-1} and the
constraints \eqref{eq:tau-1} and \eqref{eq:Ga-1} directly, then
use the results to reconstruct the generating function.

One important input for the flux attractor equations is the prepotential,
which determines the $F_{I}$ and $F_{IJ}$ via\begin{eqnarray*}
F_{I}=\partial_{I}F\,, &  & F_{IJ}=\partial_{I}\partial_{J}F\,.\end{eqnarray*}
In our first example we will study a particular $\mathbb{Z}_{4}$
orbifold of $T^{6},$ which gives rise to a prepotential 
\begin{equation}
F_{T^{6}/\mathbb{Z}_{4}}=-iZ^{0}Z^{1}\,.\label{eq:Prepotential-Z4}
\end{equation}
In the second example, we will use the STU prepotential,
\begin{equation}
F_{\mathrm{STU}}=\frac{Z^{1}Z^{2}Z^{3}}{Z^{0}}\,.\label{eq:Prepotential-STU}
\end{equation}
The simplicity of the $T^{6}/\mathbb{Z}_{4}$ example makes it easy
to demonstrate the logic of both the flux attractor equations and
the generating function. While the STU example is more involved, we
believe it is representative of what one would find when studying
the large class of cubic prepotentials.

An interesting property of the attractor equations \eqref{eq:m-1},
\eqref{eq:e-1} and constraints \eqref{eq:tau-1}, \eqref{eq:Ga-1},
is that they do not include or require detailed information about
the space of hypermoduli. While we might have imagined that e.g. the
triple intersection numbers would play an important role, at least
in the $H_{a}=0$ equations, they do not. Rather, we only need to
know $h_{-}^{\left(1,1\right)},$ which determines the number of different
geometric fluxes $r_{a}$ that can induce new F-terms, and thus stabilize
additional moduli. In the following we will carefully establish that
there are constructions that give rise to these prepotentials that
\emph{also} have $h_{-}^{\left(1,1\right)}\neq0.$

\subsection{$T^{6}/\mathbb{Z}_{4}$}

In this example, a relatively simple prepotential will allow us to
compute the generating function $\mathcal{G}$ for generic fluxes.
After describing the orbifold construction that gives rise to \eqref{eq:Prepotential-Z4},
we will solve the attractor equations \eqref{eq:m-1} and \eqref{eq:e-1}.
This gives the potentials $\phi^{I}$ and $\theta_{I}$ as functions
of the complex fluxes, which we will then use to write the generating
function $\mathcal{G}.$ We will also write the system of equations
that determines the values of the stabilized moduli as functions of
the real fluxes, and see that two hypermoduli can be stabilized.

\subsubsection{Orbifold Construction}

Let us consider an $\mathcal{N}=2$ supersymmetric orbifold $T^{6}/\Z_{4}$,
where the action of $\Z_{4}$ on the complex coordinates is generated
by 
\begin{equation}
\Theta\cdot\left(z_{1},z_{2},z_{3}\right)=\left(iz_{1},iz_{2},-z_{3}\right).
\end{equation}
 The untwisted sector of this orbifold gives rise to 5 $(1,1)$-forms
and one $(2,1)$-form. The twisted sector content depends on which
$T^{6}$ lattice we are acting. To be concrete, let us pick the $A_{3}^{2}$
lattice (the root lattice of $SU(4)\times SU(4)$). For this choice
the twisted sectors contribute 20 $\left(1,1\right)$-forms but no
3-forms (see, e.g. \citep{Reffert:2007im}), so the only complex structure
moduli will come from the untwisted sector, and we do not need to
perform any truncations when computing the prepotential or, eventually,
the generating function.

Now let us construct an $\mathcal{N}=1$ supersymmetric orientifold
by combining the involution 
\begin{equation}
\s\cdot\left(z_{1},z_{2},z_{3}\right)=\left(z_{1},-z_{2},z_{3}\right),
\end{equation}
 with $\Om(-1)^{F_{L}}$, where $\Om$ here represents a worldsheet
parity transformation and $F_{L}$ is the left-moving fermion number
on the worldsheet. The involutions $\s$ and $\Theta^{2}\s$ each
give rise to sets of untwisted sector O7-planes, while the involutions
$\Theta\s$ and $\Theta^{3}\s$ give rise to twisted sector O7-planes
which wrap exceptional divisors at the $\t^{2}$ fixed points. There
are no O3-planes in this model.

Under this orientifold involution, three of the untwisted sector $\left(1,1\right)$-forms
are invariant, while the other two change sign (all of the twisted
sector $\left(1,1\right)$-forms are invariant), giving $h_{-}^{\left(1,1\right)}=2$
and $h_{+}^{\left(1,1\right)}=23.$ All of the 3-forms change sign,
so $h_{-}^{\left(2,1\right)}=1$ and $h_{+}^{\left(2,1\right)}=0.$
Thus, in principle we can turn on geometric fluxes $r_{1}$ and $r_{2}$
as well as $H_{3}$ and $F_{3}$, and each of these 3-forms has four
components.

For a certain choice of symplectic basis, the coefficients of the
holomorphic three form \eqref{eq:Omegaexpa} correspond to a prepotential

\begin{equation}
F=-iZ^{0}Z^{1}.\label{eq:Z4Prepotential}
\end{equation}
With this information we can turn to a computation of the generating
function.

\subsubsection{Solutions for Potentials and $\mathcal{G}$}

Using \eqref{eq:NewMag-1}-\eqref{eq:NewElec-2} and the prepotential
(\ref{eq:Z4Prepotential}), we can solve for $e_{I}$ and $\t_{I}$
in terms of $m^{I}$ and $\phi^{I}$. 
\begin{eqnarray}
e_{0} & = & \overline{CF_{0}}+L^{J}F_{0J}=i\left(\overline{CZ^{1}}-L^{1}\right)=i\phi^{1},\\
e_{1} & = & \overline{CF_{1}}+L^{J}F_{1J}=i\left(\overline{CZ^{0}}-L^{0}\right)=i\phi^{0},
\end{eqnarray}
 which inverts to 
\begin{equation}
\phi^{0}=-ie_{1},\qquad\phi^{1}=-ie_{0}.
\end{equation}
 Similarly, we have 
\begin{equation}
\t_{0}=im^{1},\qquad\t_{1}=im^{0}.
\end{equation}
Inserting these results into the expression (\ref{eq:Gen-1b}) we
find 
\begin{equation}
\mathcal{G}=-\frac{i}{\tau-\bar{\tau}}\left(e_{0}\overline{e_{1}}+e_{1}\overline{e_{0}}+m^{0}\overline{m^{1}}+m^{1}\overline{m^{0}}\right).
\end{equation}

\subsubsection{Solutions for Hypermoduli and $\widetilde{\mathcal{G}}$}

We can now derive the constraints. From the complex conjugate of \eqref{eq:G-Ga-1}
we find 
\begin{equation}
\left.\frac{\p\mathcal{G}}{\p\overline{G^{a}}}\right|_{\R}=\frac{i}{\tau-\bar{\tau}}\left(r_{a1}e_{0}+r_{a0}e_{1}+r_{a}^{1}m^{0}+r_{a}^{0}m^{1}\right)=0\,,
\end{equation}
and after imposing this constraint we can write the complex conjugate
of \eqref{eq:G-tau-3} as 
\begin{equation}
\left.\frac{\p\mathcal{G}}{\p\bar{\tau}}\right|_{\R}=-\frac{2i}{\left(\tau-\bar{\tau}\right)^{2}}\left(e_{0}e_{1}+m^{0}m^{1}\right)=0\,.
\end{equation}

Setting these expressions to zero will stabilize some of our hypermoduli.
Now all three of $H_{3}$, $r_{1}$, and $r_{2}$ must be symplectically
orthogonal by the tadpole conditions \eqref{eq:NS-Tadpole}, but in
our model a symplectically orthogonal set of 3-forms is at most two-dimensional.
Because of this, we can only hope to fix at most two linear combinations
of the three moduli $\tau$, $G^{1}$, and $G^{2}$. More explicitly,
if the two independent orthogonal three-forms are denoted $\xi_{1}$
and $\xi_{2}$, and we write $H_{3}=A_{\tau}\xi_{1}+B_{\tau}\xi_{2}$,
$r_{a}=A_{a}\xi_{1}+B_{a}\xi_{2}$, then the complex flux is given
by $G_{3}=F_{3}-x_{1}\xi_{1}-x_{2}\xi_{2}$, where $x_{1}=A_{\tau}\tau+A_{a}G^{a}$
and $x_{2}=B_{\tau}\tau+B_{a}G^{a}$. Since the minimization procedure
depends on the hypermoduli only via the complex flux, we can only
hope to stabilize the linear combinations $x_{1}$ and $x_{2}$, leaving
a third linear combination unfixed.


For completeness, we present the reduced generating function for this model.  Recalling that the tadpole conditions should not be enforced until the end of the calculation, straightforward (but lengthy) algebraic manipulations lead us to $\widetilde{\mathcal{G}}$.  To simplify the expression, it is convenient to define an inner product on three-forms,
\begin{multline}
\left\langle x^0\alpha_0+x^1\alpha_1-x_0\beta^0-x_1\beta^1,y^0\alpha_0+y^1\alpha_1-y_0\beta^0-y_1\beta^1\right\rangle \\
=x^0y^1+x^1y^0+x_0y_1+x_1y_0.
\end{multline}

We then have (for generic fluxes)
\begin{eqnarray}
\widetilde{\mathcal{G}} &=& \pm\frac{1}{\left\langle r_1,r_1\right\rangle\left\langle r_2,r_2\right\rangle-\left\langle r_1,r_2\right\rangle^2}\left[\left(\left\langle r_1,r_1\right\rangle\left\langle r_2,r_2\right\rangle\left\langle F_3,F_3\right\rangle-\left\langle r_1,r_2\right\rangle^2\left\langle F_3,F_3\right\rangle\right.\right.\nonumber\\
&& \qquad\left.\left.-\left\langle r_1,r_1\right\rangle\left\langle r_2,F_3\right\rangle^2-\left\langle r_2,r_2\right\rangle\left\langle r_1,F_3\right\rangle^2+2\left\langle r_1,r_2\right\rangle\left\langle r_1,F_3\right\rangle\left\langle r_2,F_3\right\rangle\right)\right.\nonumber\\
&& \left.\times\left(\left\langle r_1,r_1\right\rangle\left\langle r_2,r_2\right\rangle\left\langle H_3,H_3\right\rangle-\left\langle r_1,r_2\right\rangle^2\left\langle H_3,H_3\right\rangle-\left\langle r_1,r_1\right\rangle\left\langle r_2,H_3\right\rangle^2\right.\right.\nonumber\\
&& \qquad\left.\left.-\left\langle r_2,r_2\right\rangle\left\langle r_1,H_3\right\rangle^2+2\left\langle r_1,r_2\right\rangle\left\langle r_1,H_3\right\rangle\left\langle r_2,H_3\right\rangle\right)\right.\nonumber\\
&& \left.-\left(\left\langle r_1,r_1\right\rangle\left\langle r_2,r_2\right\rangle\left\langle F_3,H_3\right\rangle-\left\langle r_1,r_2\right\rangle^2\left\langle F_3,H_3\right\rangle\right.\right.\nonumber\\
&& \qquad\left.\left.-\left\langle r_1,r_1\right\rangle\left\langle r_2,F_3\right\rangle\left\langle r_2,H_3\right\rangle-\left\langle r_2,r_2\right\rangle\left\langle r_1,F_3\right\rangle\left\langle r_1,H_3\right\rangle\right.\right.\nonumber\\
&& \qquad\left.\left.+\left\langle r_1,r_2\right\rangle\left\langle r_1,F_3\right\rangle\left\langle r_2,H_3\right\rangle+\left\langle r_1,r_2\right\rangle\left\langle r_2,F_3\right\rangle\left\langle r_1,H_3\right\rangle\right)^2\right]^{1/2},
\end{eqnarray}
where the plus sign is taken if 
\begin{multline}
\left\langle r_1,r_1\right\rangle\left\langle r_2,r_2\right\rangle\left\langle H_3,H_3\right\rangle-\left\langle r_1,r_2\right\rangle^2\left\langle H_3,H_3\right\rangle-\left\langle r_1,r_1\right\rangle\left\langle r_2,H_3\right\rangle^2 \\
-\left\langle r_2,r_2\right\rangle\left\langle r_1,H_3\right\rangle^2+2\left\langle r_1,r_2\right\rangle\left\langle r_1,H_3\right\rangle\left\langle r_2,H_3\right\rangle<0,
\end{multline}
and the minus sign is taken otherwise.

\subsection{The STU Model}

With this example we add geometric fluxes to a compactification with
an STU prepotential. This example was studied carefully in the absence
of geometric fluxes in \citep{Larsen:2009fw,Duff:1995sm,Behrndt:1996hu,LopesCardoso:1996yk,Behrndt:1996jn}.
Substituting the symplectic section determined by \eqref{eq:Prepotential-STU}
into \eqref{eq:e-1}, the electric attractor equations become
\begin{eqnarray}
e_{0} & = & -\frac{\overline{CZ}^{1}\overline{CZ}^{2}\overline{CZ}^{3}}{\left(\overline{CZ}^{0}\right)^{2}}+2L^{0}\frac{CZ^{1}CZ^{2}CZ^{3}}{\left(CZ^{0}\right)^{3}}-\left(L^{1}\frac{CZ^{2}CZ^{3}}{\left(CZ^{0}\right)^{2}}+\mathrm{cyc.}\right),\label{eq:e0-STU}\\
e_{1} & = & -\frac{\overline{CZ}^{2}\overline{CZ}^{3}}{\overline{CZ}^{0}}-L^{0}\frac{CZ^{2}CZ^{3}}{\left(CZ^{0}\right)^{2}}+L^{2}\frac{CZ^{3}}{CZ^{0}}+L^{3}\frac{CZ^{2}}{CZ^{0}}\,.\label{eq:e1-STU}
\end{eqnarray}
Cyclic permutations of \eqref{eq:e1-STU} give the remaining two electric
attractor equations. Since we can use \eqref{eq:m-1} to rewrite the
$L^{I}$ in terms of $m^{I}$ and $CZ^{I},$ these are four complex,
non-holomorphic, non-linear equations for the $CZ^{I}.$ By using
\eqref{eq:NewMag-1} we could recast \eqref{eq:e0-STU} and \eqref{eq:e1-STU}
as equations for the $\phi^{I}$ rather than the $CZ^{I}$. Rather
than solve directly for the $CZ^{I}$ or the potentials, we will find
it most useful to use $z^{i}=Z^{i}/Z^{0},$ so that \eqref{eq:e0-STU}
and \eqref{eq:e1-STU} are considered as equations for the $z^{i}$
and $CZ^{0}.$ 

While black hole attractor equations with the STU prepotential can
be solved explicitly for arbitrary black hole charges, the attractor
equations \eqref{eq:e0-STU} and \eqref{eq:e1-STU} do not admit an
explicit solution for general fluxes. Since we are interested in finding
explicit solutions that illuminate the results of sections \eqref{sec:GeomAttractorEqs}
and \eqref{sec:GenFnsGeom}, we will only turn on four components
of $F_{3},$ four components of $H_{3},$ and six geometric fluxes.
These 14 real flux components will allow us to explicitly stabilize
the three complex vector moduli, $z^{i}$, and four complex hypermoduli,
$\tau$ and three of the $G^{a}.$ While it is not possible to compute
explicitly the generating function $\mathcal{G}$ for these fluxes,
we will compute the \emph{reduced} generating function $\widetilde{\mathcal{G}},$
with the result given in equation \eqref{eq:G-OffShell}.

An issue that will arise at several points in our analysis is the
role of the tadpole constraints \eqref{eq:NS-Tadpole}. Once we are
analyzing equations involving real fluxes, imposing the tadpole constraints
will consistently lead to significantly simpler expressions for the
stabilized values of the moduli, the mass parameters, and the generating
function. While simplifying with the tadpole constraints will not
alter the algebraic relationships between these quantities, they do
affect their \emph{derivatives.} Since one of our goals is to illustrate
how derivatives of the generating function reproduce the moduli and
mass parameters, the primary results of sections \ref{sub:RealVectors}-\ref{sub:RealGen}
will be presented both with and without the tadpole constraints \eqref{eq:NS-Tadpole}
imposed.

\subsubsection{The Enriques Calabi-Yau and the STU Prepotential}

We saw in section \eqref{sub:hab} that geometric fluxes could only
induce new F-terms when $h_{-}^{\left(1,1\right)}\neq0.$ Unfortunately,
the standard orbifold construction that leads to the STU prepotential,
$T^{6}/\mathbb{Z}_{2}\times\mathbb{Z}_{2},$ has $h_{-}^{\left(1,1\right)}=0.$
Another construction that leads to the STU prepotential, but which
has $h_{-}^{\left(1,1\right)}=8$ is an orientifold of the Enriques
Calabi-Yau.

The construction of the Enriques Calabi-Yau \citep{Ferrara:1995yx}
begins with $K3\times T^{2}$. The \textbf{$K3$ }factor admits the
freely-acting Enriques involution, $\theta_{1},$ under which the
holomorphic 2-form is odd.\textbf{ }Orbifolding $K3$ by $\theta_{1}$
would give the Enriques surface, but we will instead orbifold $K3\times T^{2}$
by $\theta_{1}\theta_{2},$ where $\theta_{2}$ takes the torus coordinate
$z^{3}$ to $-z^{3}.$ The resulting surface is a self-mirror Calabi-Yau
with $h^{\left(1,1\right)}=h^{\left(2,1\right)}=11.$ In the orbifold
limit of the underlying $K3$ factor, the untwisted sector contributes
$h^{\left(1,1\right)}=h^{\left(2,1\right)}=3,$ while the twisted
sector contributes $h^{\left(1,1\right)}=h^{\left(2,1\right)}=8.$
The prepotential is governed by the triple intersection numbers 
\begin{eqnarray}
\kappa_{123} & = & 1\,,\label{eq:k123}\\
\kappa_{3ab} & = & C_{ab}\,,
\end{eqnarray}
where $C_{ab}$ is the Cartan matrix of $E_{8},$ and $a,b=4,...,11.$
Type II compactifications on the Enriques Calabi-Yau have $\mathcal{N}=2$
supersymmetry.

The final step in the construction is the orientifold projection~\citep{Grimm:2007xm},
which reduces the amount of supersymmetry to $\mathcal{N}=1.$ This
employs a second involution which gives $-1$ when acting on the 2-forms
$\omega_{a},$ $+1$ when acting on $\omega_{1}$ and $\omega_{2},$
and inverts the $T^{2}.$ This splits the 2-form cohomology such that
$h_{-}^{\left(1,1\right)}=8$ and $h_{+}^{\left(1,1\right)}=3.$ Because
the 3-forms are constructed by wedging together 2-forms on the underlying
$K3$ with 1-forms on the underlying $T^{2},$ the 3-form cohomology
splits with $h_{-}^{\left(2,1\right)}=3$ and $h_{+}^{\left(2,1\right)}=8.$
The triple intersection numbers \eqref{eq:k123} determine that three
surviving complex structure moduli will be governed by the STU prepotential.

\subsubsection{Complex Fluxes and the Vector Moduli}

Since we cannot explicitly solve the attractor equations \eqref{eq:e0-STU}
and \eqref{eq:e1-STU} with generic fluxes, we impose the following
reality conditions on the complex fluxes:
\begin{eqnarray}
\overline{m}^{0} & = & m^{0}\,,\label{eq:mo-real}\\
\overline{m}^{i} & = & -m^{i}\,,\\
\overline{e}_{0} & = & -e_{0}\,,\\
\overline{e}_{i} & = & e_{i}\,.\label{eq:ei-real}
\end{eqnarray}
We also make a complementary ansatz for the potentials:
\begin{eqnarray}
\overline{\phi}^{0} & = & \phi^{0}\,,\\
\overline{\phi}^{i} & = & -\phi^{i}\,,\\
\overline{\theta}_{0} & = & -\theta_{0}\,,\\
\overline{\theta}_{i} & = & \theta_{i}\,.
\end{eqnarray}
This reduction was previously utilized in \citep{Larsen:2009fw},
where it was found to be a useful compromise between completely general
fluxes (where \eqref{eq:e0-STU} and \eqref{eq:e1-STU} cannot be
solved explicitly) and solubility (since overly simple fluxes do not
stabilize all of the moduli). 

An important feature of the attractor equations \eqref{eq:m-1} and
\eqref{eq:e-1} is that the fluxes enter only via the complex fluxes
$m^{I}$ and $e_{I}.$ This means that they lead to the same solutions
for the moduli and mass parameters as functions of the complex fluxes,
with or without geometric fluxes. Since \eqref{eq:e0-STU} and \eqref{eq:e1-STU}
were already solved in \citep{Larsen:2009fw}, we simply quote the
solutions:
\begin{eqnarray}
CZ^{0} & = & \frac{1}{4}\left(m^{0}-i\sum_{i}m^{i}\sqrt{\frac{m^{0}e_{i}}{e_{0}m^{i}}}\right),\label{eq:CZ0-STU}\\
z^{i} & = & -i\sqrt{\frac{e_{0}m^{i}}{m^{0}e_{i}}}\,,\label{eq:zi-STU}
\end{eqnarray}
with no summation over $i$ in \eqref{eq:zi-STU}. The requirement
that the metric on moduli space remain positive, which in turn requires
$\mbox{Im}\left(z^{i}\right)<0$ and $\mbox{Im}\left(\tau\right)>0,$
implies a condition on the complex fluxes: 
\begin{equation}
i\frac{m^{I}}{e_{I}}>0\,,\label{eq:sign-STU}
\end{equation}
with no summation over $I$. This implies that the quantities under
the square roots in \eqref{eq:CZ0-STU} and \eqref{eq:zi-STU} are
real and positive, and we will ensure for the remainder of this section
that only real, positive quantities appear under square roots. We
will also take the positive branch of all square roots.

The universal constraint \eqref{eq:tau-1} is also written in terms
of the complex fluxes alone, and so is the same with or without geometric
fluxes. Generically, it takes the form of a condition that the complex
fluxes must satisfy. We again quote the result from \citep{Larsen:2009fw}:
\begin{equation}
\frac{e_{0}m^{1}m^{2}m^{3}}{m^{0}e_{1}e_{2}e_{3}}=-1\,.\label{eq:constraint-STU}
\end{equation}
In fact this condition was used in the derivation of \eqref{eq:CZ0-STU}
and \eqref{eq:zi-STU}, where it helped to find compact and explicit
solutions. Because of this, \eqref{eq:CZ0-STU} and \eqref{eq:zi-STU}
actually satisfy \eqref{eq:e0-STU} and \eqref{eq:e1-STU} only up
to terms that vanish after the application of \eqref{eq:constraint-STU}.

For completeness, we also record the potentials $\phi^{I}$ and $\theta_{I}.$
These are determined by substituting the solutions \eqref{eq:CZ0-STU}
and \eqref{eq:zi-STU} into \eqref{eq:NewMag-1} and \eqref{eq:NewElec-1},
which gives:
\begin{eqnarray}
\phi^{0} & = & -\frac{1}{2}\left(m^{0}+i\sum_{i}m^{i}\sqrt{\frac{m^{0}e_{i}}{e_{0}m^{i}}}\right),\label{eq:phi0-STU}\\
\phi^{1} & = & \frac{1}{2}\left(-m^{1}+i\sqrt{\frac{e_{0}m^{1}}{m^{0}e_{1}}}m^{0}+\sqrt{\frac{e_{2}m^{1}}{m^{2}e_{1}}}m^{2}+\sqrt{\frac{e_{3}m^{1}}{m^{3}e_{1}}}m^{3}\right),\label{eq:phi1-STU}\\
\theta_{0} & = & \frac{1}{2}\left(e_{0}-i\frac{e_{0}}{m^{0}}\sum_{i}m^{i}\sqrt{\frac{m^{0}e_{i}}{e_{0}m^{i}}}\right),\label{eq:theta0-STU}\\
\theta_{1} & = & \frac{1}{2}\left(ie_{0}\sqrt{\frac{m^{0}e_{1}}{e_{0}m^{1}}}-e_{1}+im^{2}\sqrt{-\frac{e_{2}e_{1}}{m^{2}m^{1}}}+im^{3}\sqrt{-\frac{e_{1}e_{3}}{m^{1}m^{3}}}\right).\label{eq:theta1-STU}
\end{eqnarray}
The expressions for $\phi^{2},$ $\phi^{3},$ $\theta_{2},$ and $\theta_{3}$
follow from cyclic permutations of \eqref{eq:phi1-STU} and \eqref{eq:theta1-STU}.
This completes our discussion of the attractor equations in terms
of complex fluxes. In order to proceed further, we will need to specify
precisely which real fluxes we are turning on.

\subsubsection{\label{sub:RealVectors}Stabilization of the Vector Moduli}

We now choose specific real fluxes consistent with the reality conditions
\eqref{eq:mo-real}-\eqref{eq:ei-real}. This will allow us to compute
the quantities associated with the vector moduli in terms of real
fluxes alone. We will also translate the sign restrictions \eqref{eq:sign-STU}
into restrictions on the real fluxes.

For $m^{0}$ and $e_{i},$ which must be real, we turn on only $m_{f}^{0}$
and $e_{i}^{f}.$ For the purely imaginary fluxes $e_{0}$ and $m^{i},$we
turn on $e_{0}^{h}$ and $m_{h}^{i},$ as well as several geometric
fluxes. In accord with the argument in section \ref{sub:GeneralCompactifications},
we turn on only three of the eight possible $r_{a},$ since we expect
that turning on more $r_{a}$ would make the constraints \eqref{eq:Ga-1}
insoluble. We will replace the $a$ index with $\tilde{i}=1,2,3,$
and turn on $r_{\tilde{i}0}$ and $r_{\tilde{1}}^{1},$ $r_{\tilde{2}}^{2},$
$r_{\tilde{3}}^{3}.$ The six real components of the geometric fluxes
are chosen so that the tadpole constraints $\int r_{\tilde{i}}\wedge r_{\tilde{j}}=0$
and $\int r_{\tilde{i}}\wedge H_{3}=0$ are automatically satisfied.
The non-trivial tadpole constraints are
\begin{equation}
0=\int r_{\tilde{i}}\wedge F_{3}=m_{f}^{0}r_{\tilde{i}0}-e_{i}^{f}r_{\tilde{i}}^{i}\,,\label{eq:g-tadpole-STU}
\end{equation}
and
\begin{equation}
n=\int F_{3}\wedge H_{3}=-m_{f}^{0}e_{0}^{h}+m_{h}^{i}e_{i}^{f}\,,\label{eq:t-tadpole-STU}
\end{equation}
where the integer $n$ is determined by the number of O3 planes and
D3 branes. 

We now write out explicitly the final set of constraints \eqref{eq:Ga-1}:
\begin{eqnarray}
0 & = & \int r_{\tilde{i}}\wedge\Omega_{3}\,.
\end{eqnarray}
For $r_{\tilde{1}}$ this reduces to
\begin{equation}
r_{\tilde{1}0}=r_{\tilde{1}}^{1}z^{2}z^{3}\,,\label{eq:geo-STU-1}
\end{equation}
with the other equations following by cyclic permutations. Inside
the Kähler cone $\mbox{Im}\left(z^{i}\right)<0,$ so we deduce that
\begin{equation}
\frac{r_{\tilde{1}0}}{r_{\tilde{1}}^{1}}<0\,.\label{eq:sign-geo-STU}
\end{equation}
We can use \eqref{eq:zi-STU} and \eqref{eq:constraint-STU} to rewrite
\eqref{eq:geo-STU-1} in terms of complex fluxes:
\begin{eqnarray}
-\frac{r_{\tilde{1}0}}{r_{\tilde{1}}^{1}} & = & \sqrt{\left(\frac{e_{0}}{m^{0}}\right)^{2}\frac{m^{2}m^{3}}{e_{2}e_{3}}}\\
 & = & -i\frac{e_{1}}{m^{1}}\sqrt{-\frac{e_{2}e_{3}}{m^{2}m^{3}}}\,.\label{eq:geo-STU-2}
\end{eqnarray}
Because the hypermoduli will enter via the complex fluxes, we would
like an expression with the complex fluxes isolated and linear. By
combining \eqref{eq:geo-STU-2} and its permutations, we find 
\begin{equation}
i\frac{m^{i}}{e_{i}}=\left(\frac{r_{\tilde{i}}^{i}}{r_{\tilde{i}0}}\right)^{2}\sqrt{-\frac{r_{\tilde{1}0}r_{\tilde{2}0}r_{\tilde{3}0}}{r_{\tilde{1}}^{1}r_{\tilde{2}}^{2}r_{\tilde{3}}^{3}}}\,,\label{eq:geo-STU-3}
\end{equation}
with no summation over $i.$ Substituting this back into \eqref{eq:constraint-STU},
we find
\begin{equation}
-i\frac{e_{0}}{m^{0}}=\sqrt{-\frac{r_{\tilde{1}0}r_{\tilde{2}0}r_{\tilde{3}0}}{r_{\tilde{1}}^{1}r_{\tilde{2}}^{2}r_{\tilde{3}}^{3}}}\,.\label{eq:geo-STU-4}
\end{equation}

For the set of geometric and 3-form fluxes we turn on $e_{i}=e_{i}^{f}$
and $m^{0}=m_{f}^{0}.$ We can therefore use \eqref{eq:geo-STU-3}
and \eqref{eq:geo-STU-4} to determine the remaining complex fluxes $m^{i}$
and $e_{0},$ which implicitly depend on the hypermoduli, in terms
of the real fluxes alone. Then \eqref{eq:CZ0-STU} and \eqref{eq:zi-STU}
give explicit expressions for the stabilized moduli and $CZ^{0}$
in terms of real fluxes:
\begin{eqnarray}
CZ^{0} & = & \frac{1}{4}\left(m_{f}^{0}+\sum_{i}e_{i}^{f}\frac{r_{\tilde{i}}^{i}}{r_{\tilde{i}0}}\right)\,,\\
z^{i} & = & i\frac{r_{\tilde{i}}^{i}}{r_{\tilde{i}0}}\sqrt{-\frac{r_{\tilde{1}0}r_{\tilde{2}0}r_{\tilde{3}0}}{r_{\tilde{1}}^{1}r_{\tilde{2}}^{2}r_{\tilde{3}}^{3}}}\,.
\end{eqnarray}
The complex potentials \eqref{eq:phi0-STU}-\eqref{eq:theta1-STU}
similarly become 
\begin{eqnarray}
\phi^{0} & = & \frac{1}{2}\left(-m_{f}^{0}+\sum_{i}e_{i}^{f}\frac{r_{\tilde{i}}^{i}}{r_{\tilde{i}0}}\right),\label{eq:phi0-STU-real}\\
\phi^{1} & = & -\frac{i}{2}\frac{r_{\tilde{1}}^{1}}{r_{\tilde{1}0}}\left(m_{f}^{0}-e_{1}^{f}\frac{r_{\tilde{1}}^{1}}{r_{\tilde{1}0}}+e_{2}^{f}\frac{r_{\tilde{2}}^{2}}{r_{\tilde{2}0}}+e_{3}^{f}\frac{r_{\tilde{3}}^{3}}{r_{\tilde{3}0}}\right)\sqrt{-\frac{r_{\tilde{1}0}r_{\tilde{2}0}r_{\tilde{3}0}}{r_{\tilde{1}}^{1}r_{\tilde{2}}^{2}r_{\tilde{3}}^{3}}}\,,\label{eq:phi1-STU-real}\\
\theta_{0} & = & \frac{i}{2}\left(-m_{f}^{0}+\sum_{i}e_{i}^{f}\frac{r_{\tilde{i}}^{i}}{r_{\tilde{i}0}}\right)\sqrt{-\frac{r_{\tilde{1}0}r_{\tilde{2}0}r_{\tilde{3}0}}{r_{\tilde{1}}^{1}r_{\tilde{2}}^{2}r_{\tilde{3}}^{3}}}\,,\label{eq:theta0-STU-real}\\
\theta_{1} & = & \frac{1}{2}\frac{r_{\tilde{1}0}}{r_{\tilde{1}}^{1}}\left(m_{f}^{0}-e_{1}^{f}\frac{r_{\tilde{1}}^{1}}{r_{\tilde{1}0}}+e_{2}^{f}\frac{r_{\tilde{2}}^{2}}{r_{\tilde{2}0}}+e_{3}^{f}\frac{r_{\tilde{3}}^{3}}{r_{\tilde{3}0}}\right),\label{eq:theta1-STU-real}
\end{eqnarray}
with the other $\phi^{i}$ and $\theta_{i}$ given by cyclic permutations
of \eqref{eq:phi1-STU-real} and \eqref{eq:theta1-STU-real}. 

So far we have not utilized the tadpole constraints \eqref{eq:g-tadpole-STU}.
After imposing the tadpole constraints, we find
\begin{eqnarray}
CZ^{0} & = & m_{f}^{0}\,,\label{eq:CZ0-OnShell}\\
z^{i} & = & \frac{i}{e_{i}^{f}m_{f}^{0}}\sqrt{-m_{f}^{0}e_{1}^{f}e_{2}^{f}e_{3}^{f}}\,,\label{eq:zi-OnShell}
\end{eqnarray}
and 
\begin{eqnarray}
\phi^{0}\,= & m^{0} & =\, m_{f}^{0}\,,\label{eq:phi0-m0-STU}\\
\phi^{i}\,= & m^{i} & =\,-\frac{i}{e_{i}^{f}}\sqrt{-m_{f}^{0}e_{1}^{f}e_{2}^{f}e_{3}^{f}}\,,\\
\theta_{0}\,= & e_{0} & =\,\frac{i}{m_{f}^{0}}\sqrt{-m_{f}^{0}e_{1}^{f}e_{2}^{f}e_{3}^{f}}\,,\\
\theta_{i}\,= & e_{i} & =\, e_{i}^{f}\,.\label{eq:thetai-ei-STU}
\end{eqnarray}
If we compare \eqref{eq:phi0-m0-STU}-\eqref{eq:thetai-ei-STU} with
\eqref{eq:NewMag-2}, we find that $L^{I}=0$ for this choice of fluxes,
so that the only non-zero mass parameter is $CZ^{0}.$ This indicates
that the only mass scale is $m_{3/2}^{2}\sim\left|CZ^{0}\right|^{2}.$

\subsubsection{Stabilization of the Hypermoduli}

In \eqref{eq:mo-real}-\eqref{eq:ei-real} we chose $e_{0}$ and $m^{i}$
to be purely imaginary. This implies that $\mbox{Re}\left(\tau\right)=\mbox{Re}\left(G^{\tilde{i}}\right)=0\,,$
so we will rewrite the hypermoduli as 
\begin{eqnarray}
\tau & = & i\tau_{2}\,,\\
G^{\tilde{i}} & = & ig^{\tilde{i}}\,,
\end{eqnarray}
where $\tau_{2}$ and $g^{\tilde{i}}$ are real. 

Our expressions \eqref{eq:geo-STU-3} and \eqref{eq:geo-STU-4} for
the complex fluxes in terms of the real fluxes, along with the definitions
\eqref{eq:mI-noQ} and \eqref{eq:eI-noQ}, give a system of linear
equations that determine the hypermoduli $\tau_2$ and $g^{\tilde{i}}:$
\begin{eqnarray}
e_{0}\,= & -i\tau_{2}e_{0}^{h}-ig^{\tilde{i}}r_{\tilde{i}0} & =\, im_{f}^{0}\sqrt{-\frac{r_{\tilde{1}0}r_{\tilde{2}0}r_{\tilde{3}0}}{r_{\tilde{1}}^{1}r_{\tilde{2}}^{2}r_{\tilde{3}}^{3}}}\,,\label{eq:e0-real}\\
m^{i}\,= & -i\tau_{2}m_{h}^{i}-ig^{\tilde{i}}r_{\tilde{i}}^{i} & =\,-ie_{i}^{f}\left(\frac{r_{\tilde{i}}^{i}}{r_{\tilde{i}0}}\right)^{2}\sqrt{-\frac{r_{\tilde{1}0}r_{\tilde{2}0}r_{\tilde{3}0}}{r_{\tilde{1}}^{1}r_{\tilde{2}}^{2}r_{\tilde{3}}^{3}}}\,.\label{eq:mi-real}
\end{eqnarray}
Note that we have not yet imposed any tadpole constraints. We can
rewrite this system of equations in matrix form,
\begin{equation}
\left(\begin{array}{cccc}
e_{0}^{h} & r_{\tilde{1}0} & r_{\tilde{2}0} & r_{\tilde{3}0}\\
m_{h}^{1} & r_{\tilde{1}}^{1} & 0 & 0\\
m_{h}^{2} & 0 & r_{\tilde{2}}^{2} & 0\\
m_{h}^{3} & 0 & 0 & r_{\tilde{3}}^{3}\end{array}\right)\left(\begin{array}{c}
\tau_{2}\\
g^{\tilde{1}}\\
g^{\tilde{2}}\\
g^{\tilde{3}}\end{array}\right)=\sqrt{-\frac{r_{\tilde{1}0}r_{\tilde{2}0}r_{\tilde{3}0}}{r_{\tilde{1}}^{1}r_{\tilde{2}}^{2}r_{\tilde{3}}^{3}}}\left(\begin{array}{c}
-m_{f}^{0}\\
e_{1}^{f}\left(r_{\tilde{1}}^{1}/r_{\tilde{1}0}\right)^{2}\\
e_{2}^{f}\left(r_{\tilde{2}}^{2}/r_{\tilde{2}0}\right)^{2}\\
e_{3}^{f}\left(r_{\tilde{3}}^{3}/r_{\tilde{3}0}\right)^{2}\end{array}\right).\label{eq:HyperMatrix}
\end{equation}
We now need only invert the $4\times4$ matrix of NS fluxes in order
to determine the hypermoduli. This can be done in general, but the
result is both quite long and not particularly illuminating. We instead
quote the result with the tadpole constraints \eqref{eq:g-tadpole-STU}
and \eqref{eq:t-tadpole-STU} imposed,
\begin{eqnarray}
\tau_{2} & = & \frac{4}{n}\sqrt{-m_{f}^{0}e_{1}^{f}e_{2}^{f}e_{3}^{f}}\,,\label{eq:t-OnShell}\\
g^{\tilde{1}} & = & \frac{1}{r_{\tilde{1}}^{1}e_{1}^{f}}\left(1-\frac{4}{n}e_{1}^{f}m_{h}^{1}\right)\sqrt{-m_{f}^{0}e_{1}^{f}e_{2}^{f}e_{3}^{f}}\,,\label{eq:g1-OnShell}
\end{eqnarray}
with the expressions for \textbf{$g^{\tilde{2}}$} and $g^{\tilde{3}}$
analogous to \textbf{\eqref{eq:g1-OnShell}. }

Now that we have computed the VEVs of all the moduli, it is interesting
to see what restrictions on the moduli and the fluxes are imposed by the combination
of the tadpole constraints, \eqref{eq:g-tadpole-STU} and \eqref{eq:t-tadpole-STU},
and the requirement that we stay inside the Kähler cone, i.e. that
the Kähler metric remain positive. While the tadpole constraints are
naturally written in terms of the fluxes alone, we can use our explicit
expressions to see how the moduli are constrained. Similarly, the
Kähler cone restrictions are naturally written in terms of the moduli,
but the explicit solutions allow us to rewrite them as restrictions
on the fluxes. 

In the case without geometric fluxes, the combination of tadpole constraints
and Kähler cone restrictions is quite restrictive. For example, in
\citep{Larsen:2009fw}, where we used the same combination of $F_{3}$
and $H_{3}$ as here, but no geometric fluxes, we found that staying
inside the Kähler cone required \begin{eqnarray*}
e_{0}^{h}m_{f}^{0}<0\,, &  & e_{1}^{f}m_{h}^{1}>0\,,\\
e_{2}^{f}m_{h}^{2}>0\,, &  & e_{3}^{f}m_{h}^{3}>0\,.\end{eqnarray*}
When we compare these with the tadpole constraint \[
n=-e_{0}^{h}m_{f}^{0}+e_{i}^{f}m_{h}^{i}\,,\]
we see that each term on the right-hand side is positive, so no individual
flux can be larger than $n.$ This renders the number of distinct
choices of $\left\{ e_{0}^{h},m_{f}^{0},e_{i}^{f},m_{h}^{i}\right\} $
finite and rather small. It also keeps the string coupling $g_{s}=1/\tau_{2}$
of order 1. We will now argue that these restrictions are far less
severe when geometric fluxes are included.

The crux of our argument is that introducing geometric fluxes does
not lead to additional Kähler cone restrictions. While we still need
to ensure that $\mbox{Im}\left(z^{i}\right)<0$ and that $t>0,$ there
is apparently no such restriction on the $g^{\tilde{i}}.$ We have
already seen the restrictions imposed on the geometric fluxes by these
requirements \eqref{eq:sign-geo-STU}, and can use the tadpole constraints
\eqref{eq:g-tadpole-STU} to find a restriction on the RR fluxes:
\begin{equation}
\frac{e_{i}^{f}}{m_{f}^{0}}<0\,.\label{eq:RR-signs}
\end{equation}
We do not, however, find any restriction%
\footnote{For example, it might appear that such a restriction would arise from
\eqref{eq:geo-STU-3} or \eqref{eq:geo-STU-4}, which involve the
complex fluxes $e_{0}$ and $m^{i}$ and so implicitly involve $e_{0}^{h}$
and $m_{h}^{i}.$ If one rewrites the complex fluxes using \eqref{eq:e0-STU},
\eqref{eq:e1-STU}, \eqref{eq:g-tadpole-STU}, and \eqref{eq:t-tadpole-STU},
both \eqref{eq:geo-STU-3} and \eqref{eq:geo-STU-4} reduce to $\sqrt{-m_{f}^{0}e_{1}^{f}e_{2}^{f}e_{3}^{f}}>0,$
which is automatically satisfied whenever \eqref{eq:RR-signs} is
satisfied.%
} on the signs of $e_{0}^{h}$ or $m_{h}^{i}.$ If we choose fluxes
such that $m_{h}^{i}/e_{0}^{h}<0,$ we can get cancellations between
the terms in \eqref{eq:t-tadpole-STU}. These cancellations allow
us to choose infinite series of fluxes that satisfy all physical constraints.
In particular, we can take the RR fluxes large and, by \eqref{eq:t-OnShell},
send the string coupling $g_{s}=1/\tau_{2}$ to zero. It would be
interesting to see how perturbative and non-perturbative corrections
might modify this result.

\subsubsection{\label{sub:RealGen}The Generating Function}

We now compute the main object of interest in this paper, the generating
function for the attractor equations. Since the result is surprisingly
simple, we will first compute the numerical value of the generating
function with the tadpole constraints \eqref{eq:g-tadpole-STU} imposed.
We will next compute the reduced generating function $\widetilde{\mathcal{G}}$
without imposing the tadpole constraints, in order to check the results
of section \ref{sec:GenFnsGeom}.

With the tadpole constraints \eqref{eq:g-tadpole-STU} and \eqref{eq:t-tadpole-STU}
imposed, we find a surprisingly simple expression for the numerical
value of the generating function $\widetilde{\mathcal{G}}.$ If we
combine the expressions for the complex potentials in \eqref{eq:phi0-m0-STU}-\eqref{eq:thetai-ei-STU}
with our explicit expression for the generating function \eqref{eq:Gen-1b},
we find
\begin{equation}
\mathcal{G}=\frac{1}{\tau-\overline{\tau}}\left\{ m^{I}\overline{e}_{I}-e_{I}\overline{m}^{I}\right\} =\frac{1}{\tau-\overline{\tau}}\int\overline{G}_{3}\wedge G_{3}\,.\label{eq:GenFn-Tadpole}
\end{equation}
The NS tadpole constraints \eqref{eq:NS-Tadpole} imply that the geometric
fluxes make no contribution to the integral in \eqref{eq:GenFn-Tadpole}.
It therefore reduces to
\begin{equation}
\mathcal{G}=-\int F_{3}\wedge H_{3}=-n\,.\label{eq:G-eq-n}
\end{equation}
An analogous generating function was derived in \citep{Larsen:2009fw}
using an identical prepotential and choice of $F_{3}$ and $H_{3},$
but with no geometric fluxes:
\begin{eqnarray}
\mathcal{G}_{\mathrm{there}} & = & n-\frac{1}{2}\left[-\mbox{sgn}\left(m_{f}^{0}\right)\sqrt{-e_{0}^{h}m_{f}^{0}}+\sum_{i}\mbox{sgn}\left(m_{h}^{i}\right)\sqrt{m_{h}^{i}e_{i}^{f}}\right]^{2}\,.\label{eq:G-no-GeoFluxes}
\end{eqnarray}
This result, along with our expressions for the moduli \eqref{eq:zi-OnShell}
and \eqref{eq:t-OnShell}, indicate that the solutions without geometric
flux cannot be recovered from the solutions with geometric flux by
formally sending the geometric fluxes to zero. Instead, this limit
is discontinuous, suggesting that there is no sense in which we can
add {}``a little'' geometric flux. This is consistent with our expectation
that the geometric fluxes obey a Dirac quantization condition, just
as the fluxes $F_{3}$ and $H_{3}$ do.

Although the expression for the generating function in \eqref{eq:G-eq-n}
is quite elegant, its derivatives will not reproduce the real potentials
$\phi_{f}^{0},$ $\phi_{h}^{i},$ $\theta_{0}^{h},$ and $\theta_{i}^{f}$
because we repeatedly used the tadpole constraints \eqref{eq:g-tadpole-STU}
and \eqref{eq:t-tadpole-STU} to simplify the expression, and using
these constraints alters the \emph{derivatives }of the generating
function. We now compute $\widetilde{\mathcal{G}}$ \emph{without}
using the tadpole constraints, and verify that its derivatives correctly
reproduce the real potentials.

In order to compute both the reduced generating function and the real
potentials, we need to compute $\tau_{2}$ without using the tadpole
constraints. If we go back to \eqref{eq:HyperMatrix} and invert we
find
\begin{equation}
\tau_{2}=\frac{\Delta_{f}}{\Delta_{h}}\sqrt{-\frac{r_{\tilde{1}0}r_{\tilde{2}0}r_{\tilde{3}0}}{r_{\tilde{1}}^{1}r_{\tilde{2}}^{2}r_{\tilde{3}}^{3}}}\,,\label{eq:t-OffShell}
\end{equation}
where we introduced the combinations 
\begin{eqnarray}
\Delta_{f} & \equiv & m_{f}^{0}+\sum_{i}\frac{r_{\tilde{i}}^{i}}{r_{\tilde{i}0}}e_{i}^{f}\,,\label{eq:Df-def}\\
\Delta_{h} & \equiv & -e_{0}^{h}+\sum_{i}\frac{r_{\tilde{i}0}}{r_{\tilde{i}}^{i}}m_{h}^{i}\,,\label{eq:Dh-def}
\end{eqnarray}
which will appear quite frequently in the following. We now substitute
the expressions for the complex potentials \eqref{eq:phi0-STU-real}-\eqref{eq:theta1-STU-real},
the expressions for the complex fluxes \eqref{eq:e0-real} and \eqref{eq:mi-real},
and the value of $\tau_{2}$ \eqref{eq:t-OffShell} into \eqref{eq:Gen-1b}
to find the reduced generating function: 
\begin{eqnarray}
\widetilde{\mathcal{G}} & = & -\frac{i}{2\tau_{2}}\left[m^{I}\overline{\theta}_{I}-e_{I}\overline{\phi}^{I}\right]\\
 & = & -\frac{1}{2}\frac{\Delta_{h}}{\Delta_{f}}\left[m_{f}^{0}\left(\Delta_{f}-2m_{f}^{0}\right)+\sum_{i}\left\{ e_{i}^{f}\frac{r_{\tilde{i}}^{i}}{r_{\tilde{i}0}}\left(\Delta_{f}-2e_{i}^{f}\frac{r_{\tilde{i}}^{i}}{r_{\tilde{i}0}}\right)\right\} \right]\\
 & = & -\frac{1}{2}\Delta_{h}\left[\Delta_{f}-2\frac{\left(m_{f}^{0}\right)^{2}+\sum_{i}\left(e_{i}^{f}r_{\tilde{i}}^{i}/r_{\tilde{i}0}\right)^{2}}{\Delta_{f}}\right].\label{eq:G-OffShell}
\end{eqnarray}
This is the principal result of this example, a single function that
summarizes all aspects of the stabilized vector moduli. Comparing
\eqref{eq:G-OffShell} with \eqref{eq:G-no-GeoFluxes}, it is interesting
that \eqref{eq:G-OffShell} has two factors, one that is independent
of $F_{3}$ and one that is independent of $H_{3},$ while each term
in \eqref{eq:G-no-GeoFluxes} mixes $F_{3}$ and $H_{3}$. Upon imposing
the tadpole constraints \eqref{eq:g-tadpole-STU} we recover \eqref{eq:G-eq-n},
as expected.

We substitute \eqref{eq:phi0-STU-real}-\eqref{eq:theta1-STU-real}
and \eqref{eq:t-OffShell} into \eqref{eq:Real-Phi} and \eqref{eq:Real-Theta}
to find the real potentials:
\begin{eqnarray}
\phi_{f}^{0} & = & \frac{1}{2}\left(\Delta_{f}-2m_{f}^{0}\right),\label{eq:Real-Phi0-OffShell}\\
\phi_{h}^{i} & = & \frac{1}{2}\Delta_{h}\frac{r_{\tilde{i}}^{i}}{r_{\tilde{i}0}}\left[1-2\frac{e_{i}^{f}r_{\tilde{i}}^{i}}{\Delta_{f}r_{\tilde{i}0}}\right],\\
\theta_{0}^{h} & = & -\frac{1}{2}\Delta_{h}\left[1-2\frac{m_{f}^{0}}{\Delta_{f}}\right],\\
\theta_{i}^{f} & = & \frac{1}{2}\frac{r_{\tilde{1}0}}{r_{\tilde{1}}^{1}}\left(\Delta_{f}-2e_{i}^{f}\frac{r_{\tilde{i}}^{i}}{r_{\tilde{i}0}}\right).\label{eq:Real-Thetai-OffShell}
\end{eqnarray}
These expressions agree with the derivatives \eqref{eq:dG-phi} and
\eqref{eq:dG-theta} of the reduced generating function \eqref{eq:G-OffShell},
up to terms that vanish when the tadpole constraints \eqref{eq:g-tadpole-STU}
are imposed, in accord with the arguments of section \eqref{sec:GenFnsGeom}.\textbf{
}This validates the generating function approach to flux attractor
equations, even after the introduction of geometric fluxes.

\acknowledgments

It is a pleasure to acknowledge helpful discussions with Ibrahima
Bah, David Berman, Chris Hull, Savdeep Sethi, and Gary Shiu. FL and
DR thank the Aspen Center for Physics, where this project was initiated,
for hospitality. FL also thanks Texas A\&M University for hospitality
during part of this project. The work of FL and RO was supported by
the DOE under grant DE-FG02-95ER40899. The work of DR was supported
by the NSF under grant PHY05-05757 and by Texas A\&M University.

\appendix

\section{\label{sec:Homogeneity}Homogeneity Conditions}

We collect here several known results about the Kähler potentials
for hypermoduli in $\mathcal{N}=1$ compactifications of Type II theories.

\subsection{Homogeneity of Hypermoduli Kähler Potentials}

In this section we will recall the form of the tree-level Kähler potential
for hypermoduli, $K,$ for various $\mathcal{N}=1$ type II compactifications.
For each we will demonstrate that $K$ is independent of the real
parts of the hypermoduli, and that $e^{-K}$ is homogeneous of degree
four in the imaginary parts of the hypermoduli.

\subsubsection{IIB O3/O7}

This is the case of greatest interest in this paper. We recall
that the hypermoduli (the scalar fields which descend from the $\N=2$
hypermultiplets) consist of the axio-dilaton $\tau$, a field $G^{a}$
corresponding to each two-form $\om_{a}$ which is odd under the orientifold
involution, and a field $T_{\al}$ corresponding to each even four-form
$\widetilde{\m}^{\al}$. In terms of the real fields (the RR potentials
$C_{0}$, $C_{2}=c^{a}\om_{a}$, and $C_{4}=\rho_{\al}\widetilde{\m}^{\al}$,
the dilaton $\phi$, the $B$-field $B_{2}=u^{a}\om_{a}$, and the
Kähler form $J=v_{\al}\m^{\al}$), they are given by%
\footnote{These conventions differ in some important ways from \citep{Grimm:2004uq}.
In particular, the definition of the $v^{\al}$ differs by a dilaton
factor ($v_{\mathrm{there}}^{\al}=e^{-\phi/2}v_{\mathrm{here}}^{\al}$),
essentially the difference between string frame and Einstein frame,
and $T_{\al}$ differs by an overall numerical factor ($T_{\al}^{\mathrm{there}}=(3i/2)T_{\al}^{\mathrm{here}}$).
They adhere more closely to \citep{Benmachiche:2006df}.%
} : 
\begin{eqnarray}
\tau & = & C_{0}+ie^{-\phi},\non\label{eq:IIBFields-Before}\\
G^{a} & = & c^{a}-\tau u^{a},\label{eqa:IIBFields}\\
T_{\al} & = & \rho_{\al}-\frac{i}{2}e^{-\phi}\left(\k v^{2}\right)_{\al}-\left(\whk cu\right)_{\al}+\hlf\tau\left(\whk u^{2}\right)_{\al},\non\label{eq:IIBFields-After}
\end{eqnarray}
as follows from (\ref{eq:IIBComplexSpinor}). We have made use of
the intersection numbers defined in (\ref{eq:IntersectionNumbers})
and (\ref{eq:IntersectionAbbreviations}).

Now the Kähler potential for these fields is 
\begin{equation}
K=-4\ln\left[-i\left(\tau-\bar{\tau}\right)\right]-2\ln\left[\mathcal{V}_{6}\right],\label{eq:IIBKahlerPotential}
\end{equation}
 where the volume 
\begin{equation}
\mathcal{V}_{6}=\frac{1}{6}\int J^{3}=\frac{1}{6}\left(\k v^{3}\right),
\end{equation}
 is implicitly viewed as a function of $T_{\al}$, $\tau$, and $G^{a}$.
One then computes the Kähler metric by using the map \eqref{eq:IIBFields-Before}-\eqref{eq:IIBFields-After}
and the expression (\ref{eq:IIBKahlerPotential}) to compute the derivatives
of $K$ with respect to the complex fields (which can only be written
explicitly in terms of the real fields, since there are no general
expressions for the $v_{\al}$ in terms of the complex fields). Inverting
that Kähler metric then gives the expressions which appear in \eqref{eq:InverseMetricBegin}-\eqref{eq:InverseMetricEnd}.

We would like to understand the scaling properties of the (exponential
of the) Kähler potential when we scale the complex fields. Looking
at \eqref{eq:IIBFields-Before}-\eqref{eq:IIBFields-After}, we see
that sending $\{\tau,G^{a},T_{\al}\}\rightarrow\{\lambda\tau,\lambda G^{a},\lambda T_{\al}\}$
for some real $\lambda$ is equivalent to an action on the real fields

\begin{equation}
\left\{ C_{0},c^{a},\rho_{\al},e^{-\phi},u^{a},v_{\al}\right\} \longrightarrow\left\{ \lambda C_{0},\lambda c^{a},\lambda\rho_{\al}\lambda e^{-\phi},u^{a},v_{\al}\right\} ,
\end{equation}
 i.e. everything scales with weight one except for $u^{a}$ and $v_{\al}$.
But then it follows immediately that 
\begin{equation}
e^{-K}=2^{4}e^{-4\phi}\mathcal{V}_{6}^{2}\,,\label{eq:IIBKahlerPotentialRealFields}
\end{equation}
 is a function of the imaginary parts of the fields which is homogeneous
of degree four, from the $e^{-\phi}$ dependence.

We can also consider the simpler case with $h_{-}^{1,1}=0$, so there
are no $G^{a}$. In this case, one can separately scale $\tau$ and
the $T_{\al}$, $\{\tau,T_{\al}\}\rightarrow\{\lambda\tau,\mu T_{\al}\}$.
In terms of the real fields, this would be 
\begin{equation}
\left\{ C_{0},\rho_{\al},e^{-\phi},v^{\al}\right\} \longrightarrow\left\{ \lambda C_{0},\m\rho_{\al},\lambda e^{-\phi},\lambda^{-\hlf}\m^{\hlf}v^{\al}\right\} .
\end{equation}
 Comparing with \eqref{eq:IIBKahlerPotentialRealFields}, we see that
$e^{-K}$ is homogeneous of degree $(1,3)$ in the scalings of $(\tau,T_{\al})$.
In particular, this fact can be used to show \eqref{eq:Simple-Homogeneity}.

\subsubsection{IIA O6}

For type IIA compactifications which are orientifolds of Calabi-Yau
manifolds, and which can contain O6-planes, the hypermoduli now come
from the complex structure moduli of the space. Indeed, in general
the orientifold involution (which, in order to preserve $\mathcal{N}=1$
supersymmetry, must be an anti-holomorphic involution of the Calabi-Yau,
and must act as minus one on the volume form of the space) can act
on the holomorphic three-form as $\s\cdot\Om_{3}=e^{2i\theta}\overline{\Om_{3}}$
for some constant phase $\theta$. Also, given a symplectic basis
$a_{K}$ and $b^{K}$, we can expand 
\begin{equation}
\Om_{3}=Z^{K}a_{K}-F_{K}b^{K}.
\end{equation}
 As usual, the $F_{K}$ here can be derived from a holomorphic prepotential
$F(Z^{K})$, which depends on our choice of symplectic basis. Then~\citep{Grimm:2004ua},
the hypermoduli come from expanding 
\begin{equation}
\Om_{c}=C_{3}+2i\mathrm{Re}\left(C\Om_{3}\right),
\end{equation}
 where $C$ is a compensator field that ensures that the expression
above is invariant under Kähler transformations. If we wish to be
more explicit, it is convenient to choose a symplectic basis in which
the $a_{K}$ are even under the orientifold involution and the $b^{K}$
are odd (we can always do this since the volume form is odd) and then
we can simply expand 
\begin{equation}
\Om_{c}=2N^{K}a_{K},\qquad N^{K}=\hlf\xi^{K}+i\mathrm{Re}\left(CZ^{K}\right),
\end{equation}
 where we have also expanded $C_{3}=\xi^{K}a_{K}$.

The Kähler potential for these fields is simply \citep{Grimm:2004uq,DeWolfe:2005uu}

\begin{equation}
K=-2\ln\left[2\int\mathrm{Re}\left(C\Om_{3}\right)\w\ast\mathrm{Re}\left(C\Om_{3}\right)\right].
\end{equation}
 From this expression it is obvious that the Kähler potential depends
only on the imaginary parts of the complex fields $N^{K}$, and that

\begin{equation}
e^{-K}=\left[2\int\mathrm{Re}\left(C\Om_{3}\right)\w\ast\mathrm{Re}\left(C\Om_{3}\right)\right]^{2},
\end{equation}
 is a homogeneous function of degree four in the $\mathrm{Im}(N^{K})$.

\subsubsection{IIA and IIB, $SU\left(3\right)\times SU\left(3\right),$ $\mathcal{N}=1$}

In fact, these homogeneity properties are even more general. Both
of the examples above could have been formulated by saying that our
complex hypermoduli fields are obtained as expansion coefficients
of a formal sum of complex forms~\citep{Benmachiche:2006df}, 
\begin{equation}
\Phi_{c}=e^{-B}C_{RR}+i\mathrm{Re}\left(\Phi\right),
\end{equation}
 where $\Phi=e^{-\phi}e^{-B+iJ}$ for IIB (see (\ref{eq:IIBComplexSpinor})),
and $\Phi=C\Om_{3}$ for IIA. In both cases, the Kähler potential
is given by 
\begin{equation}
K=-2\ln\left[i\left\langle \Phi,\overline{\Phi}\right\rangle \right],
\end{equation}
 where the pairing $\langle\cdot,\cdot\rangle$ is the Mukai pairing,
defined on even and odd forms respectively as 
\begin{equation}
\left\langle \varphi,\psi\right\rangle =\left\{ \begin{matrix}\int\left(\varphi_{0}\psi_{6}-\varphi_{2}\w\psi_{4}+\varphi_{4}\w\psi_{2}-\varphi_{6}\psi_{0}\right),\\
\int\left(-\varphi_{1}\w\psi_{5}+\varphi_{3}\w\psi_{3}-\varphi_{5}\w\psi_{1}\right).\end{matrix}\right.
\end{equation}
 Again, from this formulation it is evident that $K$ depends only
on the imaginary parts of the fields, and $e^{-K}$ is homogeneous
of degree four.

This formulation is more general than the compactifications we have
been considering so far. We could easily incorporate type IIB O5/O9
models, or we could include compactifications with $SU(3)\times SU(3)$-structure
\citep{Gualtieri:2003dx,Grana:2005sn,Benmachiche:2006df,Grana:2006hr,Cassani:2007pq,Koerber:2007xk,Grana:2008yw},
which are in some sense the most general compactifications of type
II that have $\N=1$ supersymmetry in four dimensions. Typically,
these {}``spaces\textquotedbl{} are not even geometric, but nonetheless
they have the structure displayed above, so that the effective $\N=1$
supergravity in four dimensions has a Kähler potential with the given
homogeneity properties.

\subsection{Identities Implied by Homogeneity}

In the previous section we showed that the Kähler potentials for virtually
all Type II, $\mathcal{N}=1$ compactifications obey
\begin{equation}
\eta^{A}\frac{\partial}{\partial\eta^{A}}e^{-K}=4e^{-K}\,,\label{eq:homog-4}
\end{equation}
where the index $A$ runs over all of the hypermoduli, and $\eta^{A}$
indicates the imaginary parts of those moduli. We also showed that
the Kähler potential is independent of the real parts of the hypermoduli.
We will now demonstrate how the homogeneity property \eqref{eq:homog-4}
implies \eqref{eq:homog1} and \eqref{eq:homog2}, \begin{eqnarray*}
K^{A\overline{B}}\left(\partial_{A}K\right)\left(\partial_{\overline{B}}K\right) & = & 4\,,\\
K^{A\overline{B}}\left(\partial_{\overline{B}}K\right) & = & -2i\eta^{A}\,,\end{eqnarray*}
which played a central role in section \ref{sub:GeneralCompactifications}.

We begin by relating complex derivatives to $\eta^{A}$-derivatives:

\begin{equation}
\partial_{A}=\frac{1}{2}\left(\frac{\partial}{\partial\xi^{A}}-i\frac{\partial}{\partial\eta^{A}}\right).
\end{equation}
We can use this to relate complex derivatives of the Kähler potential
$K$ to $\eta^{A}$ derivatives of $e^{-K}:$
\begin{equation}
\partial_{A}K=-e^{K}\partial_{A}\left(e^{-K}\right)=\frac{i}{2}e^{K}\frac{\partial}{\partial\eta^{A}}e^{-K}\,,\label{eq:KA-1}
\end{equation}
A similar result follows for the Kähler metric $K_{A\overline{B}}.$
We have 
\begin{equation}
\partial_{A}\partial_{\overline{B}}e^{-K}=e^{-K}\left[\left(\partial_{A}K\right)\left(\partial_{\overline{B}}K\right)-\partial_{A}\partial_{\overline{B}}K\right],
\end{equation}
so 
\begin{eqnarray}
K_{A\overline{B}}\equiv\partial_{A}\partial_{\overline{B}}K & = & \left(\partial_{A}K\right)\left(\partial_{\overline{B}}K\right)-\frac{1}{4}e^{-K}\frac{\partial}{\partial\eta^{A}}\frac{\partial}{\partial\eta^{B}}e^{K}\\
 & = & \frac{1}{4}\left[e^{2K}\left(\frac{\partial}{\partial\eta^{A}}e^{-K}\right)\left(\frac{\partial}{\partial\eta^{B}}e^{-K}\right)-e^{-K}\frac{\partial}{\partial\eta^{A}}\frac{\partial}{\partial\eta^{B}}e^{K}\right].
\end{eqnarray}
In the last step we used \eqref{eq:KA-1} to write $K_{A\overline{B}}$
in terms of $\eta^{A}$ derivatives only. If we now contract $K_{A\overline{B}}$
with $\eta^{A},$ we can use \eqref{eq:homog-4}: 
\begin{eqnarray}
\eta^{A}K_{A\overline{B}} & = & \frac{1}{4}\left[4e^{K}\frac{\partial}{\partial\eta^{B}}e^{-K}-3e^{K}\frac{\partial}{\partial\eta^{B}}e^{-K}\right]\\
 & = & \frac{1}{4}e^{K}\frac{\partial}{\partial\eta^{B}}e^{-K}\\
 & = & \frac{i}{2}\partial_{\overline{B}}K\,,
\end{eqnarray}
We can now contract with the inverse metric $K^{A\overline{B}}$ to
arrive at \eqref{eq:homog2}:
\begin{equation}
K^{A\overline{B}}\partial_{\overline{B}}K=-2i\eta^{A}\,.
\end{equation}
Contracting this expression with $\partial_{A}K$ and using \eqref{eq:homog-4}
again we find: 
\begin{eqnarray}
K^{A\overline{B}}\left(\partial_{A}K\right)\left(\partial_{\overline{B}}K\right) & = & -2i\eta^{A}\partial_{A}K\\
 & = & e^{K}\eta^{A}\frac{\partial}{\partial\eta^{A}}e^{-K}\\
 & = & 4\,.
\end{eqnarray}
This is just \eqref{eq:homog1}, so we have demonstrated that \eqref{eq:homog-4}
implies \eqref{eq:homog1} and \eqref{eq:homog2}. 

\bibliographystyle{/Users/rcoconne/Documents/LyX/hunsrt-abbrv}
\bibliography{GeometricFluxes}

\end{document}